\documentclass[draftclsnofoot,12pt, onecolumn]{IEEEtran} %双栏
\ifCLASSINFOpdf
\else
\fi

\pdfoutput=1
\usepackage{cite}
\usepackage{amsmath,amssymb,amsfonts}
\usepackage{amsthm}
\usepackage{algorithmic}
\usepackage{graphicx}
\usepackage{textcomp}
\usepackage{xcolor}
\usepackage{amssymb}
\usepackage{mathrsfs}
\usepackage{stfloats}  %公式跨栏置顶
\usepackage{lipsum}
\usepackage{subeqnarray}
\usepackage{cases}

\usepackage{graphicx}
\usepackage{longtable}
\usepackage{rotating}
\usepackage{multirow}
\usepackage{diagbox}
\usepackage{xcolor}
\usepackage{colortbl}
\usepackage{etoolbox}
\usepackage{subfigure}
\usepackage{bm}
\usepackage{booktabs}
\usepackage{makecell}
\usepackage{booktabs} %表格加粗
\usepackage[colorlinks,
            linkcolor=black,
            anchorcolor=blue,
             urlcolor=blue,
            citecolor=black
            ]{hyperref}

\makeatletter
\patchcmd{\@makecaption}
  {\scshape}
  {}
  {}
  {}
\makeatother

\usepackage{array}

\makeatletter
\newcommand*\bigcdot{\mathpalette\bigcdot@{.5}}
\newcommand*\bigcdot@[2]{\mathbin{\vcenter{\hbox{\scalebox{#2}{$\m@th#1\bullet$}}}}}
\makeatother

\newtheorem{lem}{Lemma}
\newtheorem{prop}{Proposition}
\theoremstyle{remark}

\newtheorem{rem}{Remark}

\begin{document}

%\title{Bare Demo of IEEEtran.cls for \textsc{Transactions on Magnetics}}
\title{Analysis of Age of Information in Dual Updating Systems}

\author{
\IEEEauthorblockN{Zhengchuan~Chen,~\IEEEmembership{Member,~IEEE,} Dapeng~Deng,~\IEEEmembership{Student Member,~IEEE,} Howard H. Yang,~\IEEEmembership{Member,~IEEE,}
Nikolaos Pappas,~\IEEEmembership{Senior Member,~IEEE,} Limei~Hu,~\IEEEmembership{Student Member,~IEEE,} Yunjian~Jia,~\IEEEmembership{Member,~IEEE,} Min~Wang,~\IEEEmembership{Member,~IEEE,} and Tony Q. S. Quek,~\IEEEmembership{Fellow,~IEEE}}
\thanks{
Z. Chen, D. Deng, L. Hu, and Y. Jia are with the School of Microelectronics and Communication Engineering, Chongqing University, China (Emails: \{czc, ddp, lmhu, yunjian\}@cqu.edu.cn).
H. H. Yang is with the Zhejiang University/UIUC Institute, Zhejiang University, China (Email: haoyang@intl.zju.edu.cn).
N.~Pappas is with the Department of Computer and Information Science, Link\"{o}ping University, 581 83  Link\"{o}ping, Sweden (Email: nikolaos.pappas@liu.se).
M.~Wang is with the School of Optoelectronics Engineering, Chongqing University of Posts and Telecommunications, China (Email: wangm@cqupt.edu.cn). Tony Q. S. Quek is with the Information System Technology and Design Pillar, SUTD, Singapore 487372 (Email: tonyquek@sutd.edu.sg).
}
}

%\author{
%\IEEEauthorblockN{Dapeng~Deng, Zhengchuan~Chen, Yunjian~Jia, Liang~Liang, Shuyang~Fang, Min Wang}
%\thanks{D.~Deng, Z.~Chen, Y.~Jia, L.~Liang, and S.~Fang  are with the School of  Microelectronics and Communication Engineering, Chongqing University, Chongqing 400044, China (e-mail: ddp@cqu.edu.cn; czc@cqu.edu.cn; yunjian@cqu.edu.cn; liangliang@cqu.edu.cn; fsy@cqu.edu.cn;).
%
%M.~Wang is with the School of Optoelectronics Engineering, Chongqing University of Posts and Telecommunications, Chongqing 400065, China (e-mail: wangm@cqupt.edu.cn).
%}
%}

%\author{\IEEEauthorblockN{Dapeng Deng\IEEEauthorrefmark{1},
%Montgomery Scott\IEEEauthorrefmark{3}, and
%Eldon Tyrell\IEEEauthorrefmark{4},~\IEEEmembership{Fellow,~IEEE}}
%\IEEEauthorblockA{\IEEEauthorrefmark{1}School of Electrical and Computer Engineering,
%Georgia Institute of Technology, Atlanta, GA 30332 USA}
%\IEEEauthorblockA{\IEEEauthorrefmark{2}Twentieth Century Fox, Springfield, USA}
%\IEEEauthorblockA{\IEEEauthorrefmark{3}Starfleet Academy, San Francisco, CA 96678 USA}
%\IEEEauthorblockA{\IEEEauthorrefmark{4}Tyrell Inc., 123 Replicant Street, Los Angeles, CA 90210 USA}% <-this % stops an unwanted space
%\thanks{Manuscript received December 1, 2012; revised December 27, 2012.
%Corresponding author: M. Shell (email: http://www.michaelshell.org/contact.html).}}

 \maketitle
 %\vspace{-1.0cm}
\IEEEtitleabstractindextext{
\begin{abstract}
We study the average Age of Information (AoI) and peak AoI (PAoI) of a dual-queue status update system that monitors a common stochastic process. Although the double queue parallel transmission is instrumental in reducing AoI, the out of order of data arrivals also imposes a significant challenge to the performance analysis. We consider two settings: the M-M system where the service time of two servers is exponentially distributed; the M-D system in which the service time of one server is exponentially distributed and that of the other is deterministic. For the two dual-queue systems, closed-form expressions of average AoI and PAoI are derived by resorting to the graphic method and state flow graph analysis method. Our analysis reveals that compared with the single-queue system with an exponentially distributed service time, the average PAoI and the average AoI of the M-M system can be reduced by $33.3\%$ and $37.5\%$, respectively. For the M-D system, the reduction in average PAoI and the average AoI are $27.7\%$ and $39.7\%$, respectively. Numerical results show that the two dual-queue systems also outperform the M/M/2 single queue dual-server system with optimized arrival rate in terms of average AoI and PAoI.

 %The numerical results show that when the service rates of the two sensors are the same,and the same service rate,

% Compared with the single-source just-in-time update system with exponential service time distribution, the results show that the average PAoI and the average AoI of the M-M system is dropped by $33.3\%$ and $37.5\%$, respectively, and those of the M-D system can be reduced by $27.7\%$ and $39.7\%$, respectively.The numerical results show that when the service rates of the two sensors are the same,

\end{abstract}
%\vspace{-0.5cm}
\begin{IEEEkeywords}
Age of information, dual-queue, timely status update,  status sampling network.
\end{IEEEkeywords}}

%\maketitle

\IEEEdisplaynontitleabstractindextext

% \fi
\IEEEpeerreviewmaketitle

\section{Introduction}
\allowdisplaybreaks
\IEEEPARstart{W}{ith} the widespread deployment of smart devices and real-time applications in the Internet of Things (IoT), timely status information provision of an underlying physical process has become critical. For instance, in environmental detection applications, timely updates of temperature, humidity, air pressure, and other status information directly affect the correctness of users' decisions  \cite{AuD,pappas_mag}.
In wireless body sensor networks,  real-time physiological signal (e.g., blood glucose, electrocardiogram, blood pressure, etc.) monitoring guarantees timely treatment for patients \cite{wbsn}.
Besides,  the emerging field of autonomous driving necessitates collecting and distributing  status information (e.g., velocity, acceleration,
trajectory, etc.) in real-time to achieve accurate control of vehicles \cite{vehicle_2}.
In these applications, the timeliness of information is crucial to ensure the safe and stable operation of the system.

Age of Information (AoI) was proposed as a metric to measure information freshness \cite{01}. Different from delay, AoI is destination-centric,  defined as the time elapsed since the generation of the latest successfully received update at the destination \cite{ut_1, ut_2, ut_4, book}. Specifically, at time $t$, if the monitor received an update with time-stamp $u(t)$, AoI is the random process $\Delta(t)=t-u(t)$.  Average AoI, as the most commonly used metric of AoI, is defined as the time average of AoI.
In previous literature, the AoI often refers to the time average of the random process $\Delta(t)$. To avoid confusion, in this work, AoI refers to the process $\Delta(t)$, and the average AoI refers to the time average of the process $\Delta(t)$.
Another widely used metric of AoI is the Peak Age of Information (PAoI), which is used for evaluating the maximum instantaneous AoI \cite{  PAoI_R2, ref_jit}.

\subsection{Related Works}
Early works on AoI  mainly focused on analyzing the average AoI in various transmission models, where the transmission link between transmitters and receivers is abstracted as a queueing system. For example, the authors of \cite{ut_2}  analyzed the average AoI for M/M/1, M/D/1, and D/M/1 queues with infinite queue capacity.
 A general expression of AoI for the stationary distribution was derived in \cite{A_general}. Then, the average AoI for M/G/1 and G/M/1 systems was analyzed.
The average AoI and average PAoI of loss queueing models M/M/1/1, M/M/1/2, and M/M/1/2* were derived in \cite{ref_jit}. It revealed that dropping redundant packets can effectively improve the freshness of information in heavy load data flow scenarios. Apart from single-stream queues, the analysis of AoI has been extended to multi-stream queues where a series of resource scheduling algorithms and packet management strategies were proposed to optimize the AoI performance of the considered systems. For instance, the authors of \cite{pappas_icc} proposed a queue management technique that replaces the previously queued packets by the newly arrived ones to reduce the AoI of the multi-source M/M/1 queue.
The average AoI of a multi-stream M/G/1/1 system without buffer queues was investigated in \cite{policy_1} and \cite{my_tcom} under preemptive and non-preemptive strategies, respectively. The work in \cite{pappas_iotj} devised an optimal updating  strategy for a two-stream status update system, where one stream is AoI sensitive and the other is not.

The aforementioned works primarily focused on single-queue systems, in which only one server exists. Recognizing the limitation of single-server models in capturing the characteristics of multipath transmission in communication networks \cite{multi_path}, \cite{multi_server, parallel_MM11, GG} extend the analysis of AoI into multi-server/multi-queue systems. The authors in \cite{MM2_conference} investigated the average AoI of the M/M/2 queue with two servers and further studied the M/M/$\infty$ queue with an infinite number of servers \cite{MM2}.
The author of \cite{GG} extended the AoI analysis to  GI/GI/$\infty$ queue and obtained the probability distribution of AoI.
In \cite{MPTCP}, the authors  utilized the multipath TCP (MPTCP) technology in software defined network (SDN) to improve the  timeliness of information.
The average AoI of a parallel network with multiple memoryless servers was derived in \cite{parallel_c}, where newly arrived packets can preempt service when all servers are busy.
To further explore the impact of multi-queue transmissions on AoI, a dual-server short packet communication model with deterministic service time was considered in \cite{closed_form}. The results showed that the average AoI and throughput of the dual-queue system are significantly improved compared with the single-queue system.
Then, \cite{dual} investigated the AoI gain of multi-queue transmissions in a dual-sensor update system, where two sensors synchronously transmit the same information to the destination node.
 Based on the concept of AoI, the authors of \cite{Related_02} investigated the benefits of using the correlation of information from different sensors to reduce the system estimation error and sensor update rate.

\subsection{Motivation}
From the mentioned existing works, it is confirmed that: 1) Parallel transmission strategy can improve the robustness of the system;
2) Using multi-queue parallel transmission can improve the effective data arrival rate at the receiver, thus improving the freshness of the data.
Although this strategy will increase the cost of network deployment, in real-time application-oriented systems, the focus is to ensure the freshness of received information.
To this end, we are inspired to investigate the information freshness of multi-queue parallel transmission systems.
Specifically, we consider a remote status monitoring system composed of two sensors, in which the two sensors sample the same physical process and send status updates to the monitor.
Taking the different channel conditions among transmission links into account,  it is the first time to consider the case where the service time of different servers follows different distributions in the study of parallel queues. Moreover,  to take full advantage of multi-queue transmission and send more status updates to the monitor, in this system, the two sensors work independently in asynchronous mode, which differentiates from work \cite{dual}.

This model can be applied in many multi-connectivity systems and different physical layer communication technologies can be used to diversify the considered model. For instance, in intelligent agriculture systems, to acquire a comprehensive and accurate understanding of crop production, it is necessary to deploy multiple sensors to observe environmental temperature, soil moisture, and other status information  \cite{AuD}. Accurate detection and prevention is of great significance, particularly in the building fire monitoring system. To respond to all kinds of emergencies in time, consistent sensors in different locations are installed.
Other example includes wireless camera networks where multiple cameras are required to monitor the same specific scene, so as to improve the accuracy of monitoring and realize the 3D reconstruction of the monitoring scene \cite{camera}.

\subsection{Main Contributions}
The present paper considers a status update system where two sensors independently sample a common physical process and send updates to a remote monitor.
Since the sensors have different service time, the updates may arrive out of order.
We develop analytical expressions to characterize the average PAoI and average AoI of the considered system, and quantitatively analyze the AoI performance improvement of the dual-queue system compared to a single-queue system, which provide useful guidance to the design and resource allocation of practical communication systems.
The main contributions of this work are summarized as follows:
\begin{enumerate}
\item We analyze the system state of the M-M system, where the service time of  both sensors is exponentially distributed. Based on the analysis of the AoI evolution process, we characterize the state transition diagram of the M-M system. Then, we derive closed-form expressions of the average PAoI and the average AoI by leveraging tools from Markov chains. The analysis elucidates the improvement in data freshness by utilizing multi-queue parallel transmission.
\item We investigate the system state of the M-D system, where the service time of one server is exponentially distributed, and that of the other server is deterministic. By analyzing the AoI samplepath of the M-D system in one service period  under different states, we derive the analytical expressions of  average PAoI and average AoI of the M-D system.
\item Based on the theoretical results, we carry out numerical analysis to explore the performance gain of the proposed system architecture. Specifically, we find that compared with a single-queue system where the service time is exponentially distributed, the average PAoI of the proposed M-M system decreases by $33.3\%$, and the average AoI decreases by $37.5\%$. The average PAoI of the proposed M-D system is reduced to $72.3\%$ of the single-queue update system, and the average AoI is reduced to $60.3\%$. Besides, under the same service resources, the AoI performance of the proposed dual-queue system are also significantly improved compared with the M/M/2 single queue dual-server system.
%Based on the closed-form expressions of average PAoI and average AoI, we compare the AoI performance of the  M-M system and M-D system with the single-source just-in-time update system whose service time is exponentially distributed. Numerical results show that, compared with the single-source just-in-time update system, the average PAoI of the M-M system decreases by $33.3\%$, and the average AoI decreases by $37.5\%$. The average PAoI of the M-D system is reduced to $72.3\%$ of the single-source just-in-time update system, and the average AoI is reduced to $72.3\%$.
\end{enumerate}

The remainder of this paper is organized as follows. The system model and main results are presented in Section \ref{sec:model}.
The AoI of the M-M system and the M-D system are analyzed in Section \ref{sec_MM} and Section \ref{sec:MD}, respectively.
In Section \ref{section_results}, extensive simulation results and numerical results are given to validate the reliability of theoretical analysis. Comparisons of the AoI performance between M-M and M-D systems are also presented in this section. Finally, concluding remarks are provided in Section \ref{Conclusion}.

\section{System Model and Main Results}\label{sec:model}
\begin{figure}
  \centering
  \includegraphics[scale=0.8]{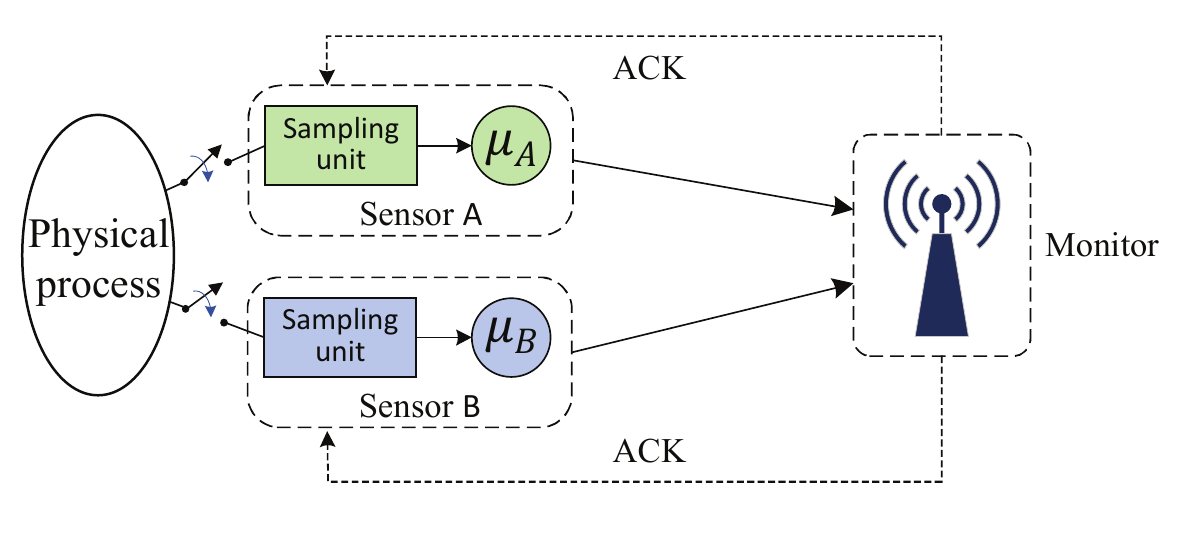}
  \caption{A dual-queue real-time monitoring system.}
  \label{system_model}
  \vspace{-1.5em}
\end{figure}

\allowdisplaybreaks
We consider a remote monitoring system consisting of two sensors and a monitor, as depicted in Fig.~\ref{system_model}.  The two sensors observe the same physical process independently and continuously  send updates to the monitor using the uplink transmission technology of  Narrowband IoT (NB-IoT), i.e., single-carrier frequency-division multiple access (SC-FDMA) \cite{NB_IOT}.
Once an update is successfully received, the monitor immediately sends an acknowledgment (ACK) to the sensor through a dedicated feedback channel.
We assume that the transmission in the feedback channel is  instantaneous  and error-free. Each sensor submits a new update once it receives an ACK, which identifies the server becomes idle.

\subsection{Multi-Queue Model}
We assume the service time $S_A$ of  sensor A follows an exponential distribution with parameter $\mu_A$. Depending on the type of service time at the redundant sensor, i.e., sensor B, we consider the following two cases respectively. \textbf{(i) M-M System}: The service time  of sensor B is exponentially distributed with parameter $\mu_B$. Note that the sensor immediately submits a new update when an ACK is received. Hence, in this case, the system can be modeled as  two parallel G/M/1/1 queues with infinite arrival rates.
 \textbf{(ii) M-D System}: The service time of sensor B is deterministic, denoted as $T$. In this case, the system can be regarded as a composition of a G/M/1/1 queue and a G/D/1/1 queue in parallel.

%Therefore, for a single source-monitor pair, it is equivalent to a M/M/1 queue whose generation rate $\lambda\to\infty$, i.e., a just-in-time update system \cite{ut_1}.

In the typical setting of AoI, the instantaneous AoI $\Delta(t)$ observed at the monitor increases linearly in time until a new update packet is received.  In this work, the monitor receives observation information from two sensors that monitor the same physical process. Due to the randomness in the service time, some updates will be obsolete when they arrive at the monitor.\footnote{An update is obsolete if its AoI is greater than that observed at the monitor, since it cannot reduce the AoI at the monitor.}
Upon receiving the obsolete updates, the server directly discard them since these updates cannot reduce the AoI observed at the monitor.
Fig.~\ref{MM_AoI_evolution} illustrates an example of the AoI evolution of sensors A and B at the monitor, where the outline of the shaded area in the figure is the real AoI evolution curve at the receiver.
 Let $Y_{(n)}$ represent the  inter-arrival time between the $(n-1)\text{th}$ and the $n\text{th}$ successfully received updates, $T_{(n)}$ denote the service time of the $n\text{th}$ successfully transmitted update, and $Q_{(n)}$ the area associated with the $n\text{th}$ update.
%The goal of this paper is to analyze the AoI of the M-M and M-D systems, and to explore the potential AoI performance gain brought by multi-queue parallel transmission.

\subsection{Summary of the Main Results}
This part summarizes the main findings of our work.
\begin{prop}\label{prop_mm_paoi}
In the M-M dual-queue parallel transmission system, the average PAoI is
\begin{equation}\label{MM_PAoI}
  \Delta_{\textnormal{M-M}}^{\textnormal{peak}}=\frac{2(\mu_A+\mu_B)}{\mu_A^2+\mu_A\mu_B+\mu_B^2}.
\end{equation}
\end{prop}
\begin{IEEEproof}
The proof is provided in Section \ref{proof_MM_paoi}.
\end{IEEEproof}

\begin{prop}\label{prop_mm_aoi}
In the M-M dual-queue parallel transmission system, the average AoI is
\begin{equation}\label{MM_AoI}
  \Delta_{\textnormal{M-M}}=\frac{2(\mu_A^2+3\mu_A\mu_B+\mu_B^2)}{(\mu_A+\mu_B)^3}.
\end{equation}
\end{prop}
\begin{IEEEproof}
The proof is provided in Section \ref{proof_MM_aoi}.
\end{IEEEproof}

According to Propositions \ref{prop_mm_paoi} and \ref{prop_mm_aoi}, we can get the following remark.
\begin{rem}
If $\mu_A=\mu_B=\mu$, the average AoI and PAoI of the considered system are $\frac{5}{4\mu}$ and $\frac{4}{3\mu}$, respectively. Compared with those of a single-queue system with service rate $\mu$ and  infinite generation rate (see \cite[Sec. \uppercase\expandafter{\romannumeral4}]{ref_jit}), both of which are $2/\mu$, the average AoI and PAoI of the M-M update system are reduced by $(1-\frac{5/(4\mu)}{2/\mu})\times100\%=37.5\%$ and $(1-\frac{4/(3\mu)}{2/\mu})\times100\%\approx33.3\%$, respectively. Please refer to Numerical results section for a more detailed discussion on the potential gains. %More detailed discussion on the potential gains are provided in numerical results section.
\end{rem}

\begin{prop}\label{prop_md_paoi}
In the M-D dual-queue parallel transmission system, the average PAoI can be expressed as
\begin{equation}\label{MD_PAoI}
  \Delta_{\textnormal{M-D}}^\textnormal{peak}=\frac{2+2\mu T+e^{\mu T}(-2+\mu T(2e^{\mu T}+\mu T))}{\mu(1+e^{\mu T}(1+e^{\mu T})\mu T)}.
\end{equation}
\end{prop}
\begin{IEEEproof}
The proof is provided in Section \ref{proof_md_paoi}.
\end{IEEEproof}

\begin{prop}\label{prop_md_aoi}
In the M-D dual-queue parallel transmission system, the average AoI can be expressed as
\begin{equation}\label{MD_AoI}
  \Delta_{\textnormal{M-D}}=\frac{3+2T\mu+e^{T\mu}(-3+(-1+2e^{T\mu})T\mu)}{T\mu^{2}e^{2T\mu}}.
\end{equation}
\end{prop}
\begin{IEEEproof}
The proof is provided in Section \ref{proof_md_aoi}.
\end{IEEEproof}

The following conclusion can be drawn from Propositions \ref{prop_md_paoi} and \ref{prop_md_aoi}.
\begin{rem}
When $T=1/\mu$, the average AoI and PAoI of the M-D system are $\frac{2e^{2}-4e+5}{\mu e^{2}}$ and $\frac{5e^{2}-e+4}{(e^2+e+1)\mu}$, respectively. Compared with the average AoI and PAoI of the single-queue update system with service rate $\mu$ and  infinite generation rate (see \cite[Sec. \uppercase\expandafter{\romannumeral4}]{ref_jit}), both of which are $2/\mu$, the average AoI and PAoI of the M-D system are reduced by $\big(1-\frac{(2e^2-4e+5)/(\mu e^{2})}{2/\mu}\big)\times100\%\approx39.7\%$ and $\big(1-\frac{(5e^{2}-e+4)/[(e^2+e+1)\mu]}{2/\mu}\big)\times100\%\approx27.7\%$, respectively.
\end{rem}

The following sections provide the detailed mathematical proofs of the findings described above.

\section{AoI of the M-M System}\label{sec_MM}
\allowdisplaybreaks
In this section, we adopt the graphic method and state transfer method to circumvent the issue of out of order packet arrivals caused by the redundant sensor and derive analytical expressions for the average AoI and PAoI for the M-M system\footnote{Note that the average AoI can be also derived by utilizing stochastic hybrid systems.}.

\begin{figure}
  \centering
  \includegraphics[scale=0.45]{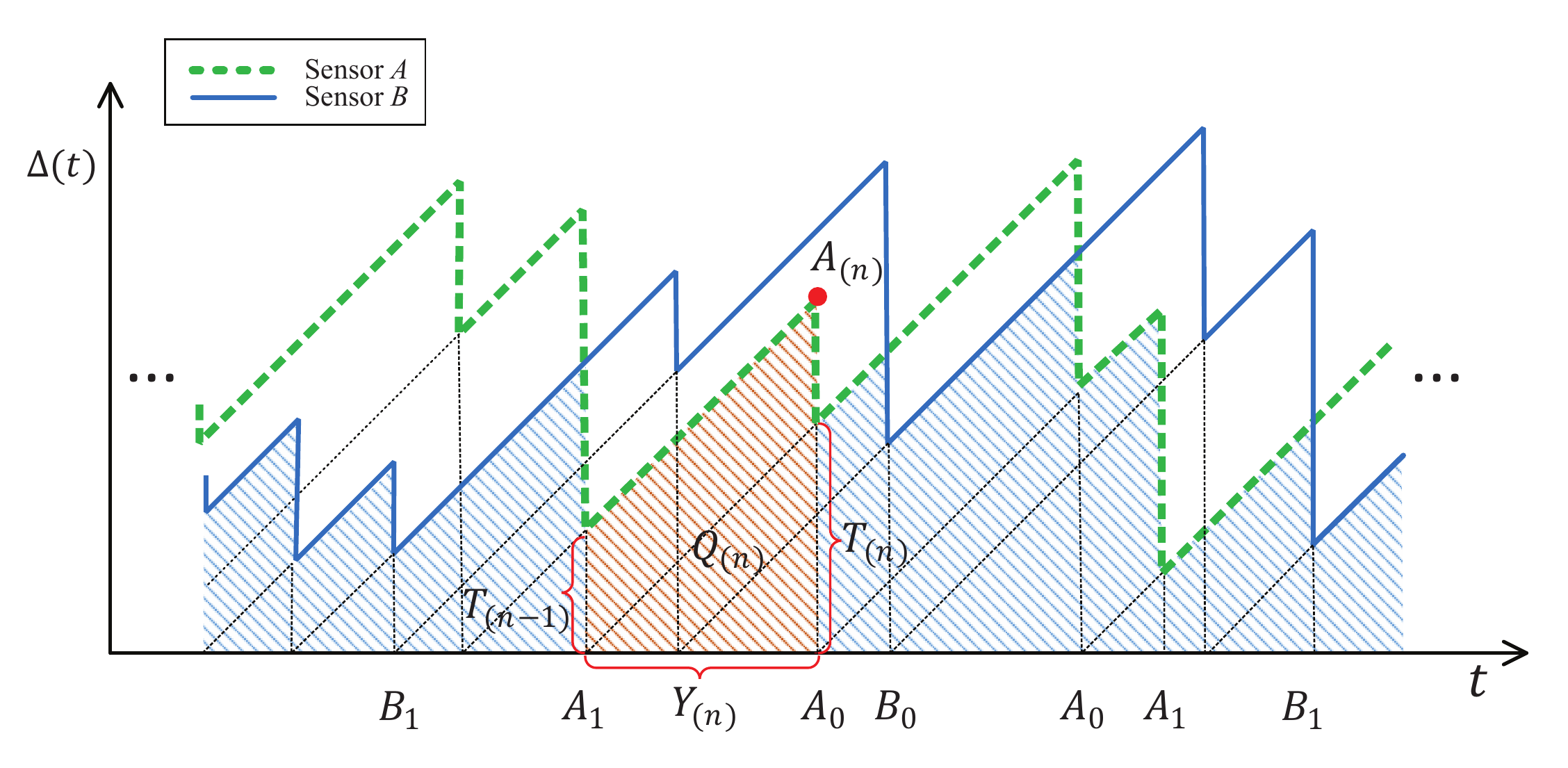}
  \caption{A sample path of AoI observed at the monitor.}
  \label{MM_AoI_evolution}
   \vspace{-1.5em}
\end{figure}

In order to analyze the time average of AoI, we define the state of the system at the time when the AoI is refreshed as four states: $A_0$, $A_1$, $B_0$, and $B_1$, as shown in Fig.~\ref{MM_AoI_evolution}.
State $A_0$ indicates that an update from sensor A is successfully received by the monitor and the generation time of this update is earlier than the update being served in sensor B. Since the update being served in sensor B is generated later than the update currently successfully sent by sensor A, the information carried by the update being served in sensor B is up-to-date and useful, and this update can reduce the AoI observed at the monitor upon reception.  State $A_1$ indicates that an update from sensor A is successfully received by the monitor and the generation time of this update is later than that of the update being served in sensor B, which means the update being served in sensor B is obsolete. Similarly, states $B_0$ and $B_1$ indicate that the monitor successfully received a fresh update from sensor B and the update being served in sensor A is fresh and stale, respectively.

\begin{figure}
  \centering
  \includegraphics[scale=0.5]{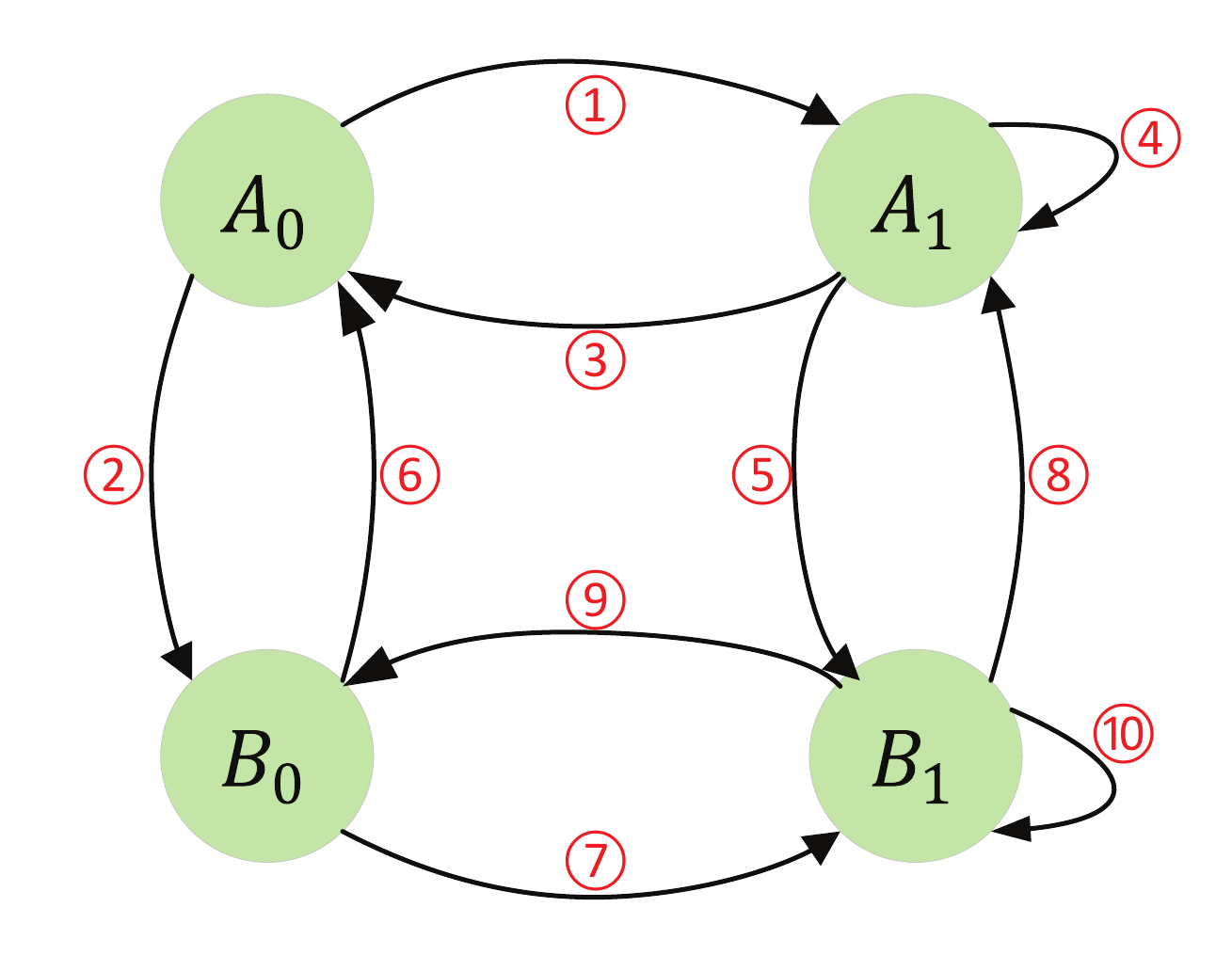}
  \caption{The state transition diagram of the M-M system.}
  \label{MM_state_transition}
  \vspace{-1.5em}
\end{figure}

The system state transition diagram can be presented as a graph ($\mathcal{Q}$, $\mathcal{L}$) shown in Fig.~\ref{MM_state_transition}, where the discrete system state $q_{l}\in\mathcal{Q}$ is transferred to state $q_{l}^{\prime}\in\mathcal{Q}$ through path $l\in\mathcal{L}$ with transition probability $p_l$. The system state space is $\mathcal{Q}=\{A_0, A_1, B_0, B_1\}$ and the space of the state transition path is $\mathcal{L}=\{1, 2, ..., 10\}$.
In addition, when the system is transferred by path $l$, we define $Y_l$ as the inter-arrival time of update at the monitor, i.e., the elapsed time for the system to transit from state $q_{l}$ to state $q_{l}^{\prime}$, and $T_l$ as the service time of the successfully received update.

\subsection{Average PAoI Analysis}\label{proof_MM_paoi}
%%Before starting to calculate the average AoI of the M-M system, let us define some necessary variables.
%Let $S_A$ and $S_B$ indicate the service time required for sensor A  and sensor B  to transmit an update, respectively.
%In the M-M system,  $S_A$ and $S_B$ follow the  exponential distribution with parameters of $\mu_A$ and $\mu_B$, respectively. The variable $S_{AA}$ represents the sum of the service times of two independent packets from sensor A, i.e., $S_{AA}$ follows the second-order Erlang distribution with parameter $\mu_A$, and its probability density function is
%\begin{equation}\label{two_Er}
%  f_{S_{AA}}(x)=\left\{
%\begin{aligned}
%&\mu_A^2xe^{-\mu_A x}&\quad&x>0, \\
%&0&\quad&x\leq0.
%\end{aligned}
%\right.
%\end{equation}
This section analyzes the average PAoI of the M-M system by establishing a Markov chain to capture the state transitions.
The relevant statistics are summarized in Table \ref{State_pro}. Below we provide the derivation for the entries in Table \ref{State_pro}.

\begin{table}
  \centering
  \caption{Statistics corresponding to each transition in Fig.~\ref{MM_state_transition}.}\label{State_pro}
  \renewcommand\arraystretch{1.5}
  \resizebox{\textwidth}{!}{
  \begin{tabular}{!{\vrule width1.2pt}c|c|c|c|c|c|c|c|c|c|c|c!{\vrule width1.2pt}}
  \Xhline{1.2pt}
  $l$ & $q_l\!\to\! q_l^{\prime}$ & $p_l$ & $\mathbb{E}[T_l]$ & $\mathbb{E}[Y_l]$ & $\mathbb{E}[Y_l^2]$ &  $l$ & $q_l\!\to\! q_l^{\prime}$ & $p_l$ & $\mathbb{E}[T_l]$ & $\mathbb{E}[Y_l]$ & $\mathbb{E}[Y_l^2]$\\
  \hline
    1 & $A_0\!\to\! A_1$ & $\frac{\mu_A}{\mu_A\!+\!\mu_B}$     & $\frac{1}{\mu_A\!+\!\mu_B}$ & $\frac{1}{\mu_A\!+\!\mu_B}$ & $\frac{2}{(\mu_A\!+\!\mu_B)^2}$ &
    6 & $B_0\!\to\! A_0$ & $\frac{\mu_A}{\mu_A\!+\!\mu_B}$     & $\frac{2}{\mu_A\!+\!\mu_B}$ & $\frac{1}{\mu_A\!+\!\mu_B}$ & $\frac{2}{(\mu_A\!+\!\mu_B)^2}$\\
  \hline
   2 & $A_0\!\to\! B_0$ & $\frac{\mu_B}{\mu_A\!+\!\mu_B}$      & $\frac{2}{\mu_A\!+\!\mu_B}$ & $\frac{1}{\mu_A\!+\!\mu_B}$ & $\frac{2}{(\mu_A\!+\!\mu_B)^2}$ &
   7 & $B_0\!\to\! B_1$ & $\frac{\mu_B}{\mu_A\!+\!\mu_B}$      & $\frac{1}{\mu_A\!+\!\mu_B}$ & $\frac{1}{\mu_A\!+\!\mu_B}$ & $\frac{2}{(\mu_A\!+\!\mu_B)^2}$\\
  \hline
   3 & $A_1\!\to\! A_0$ & $\frac{\mu_A\mu_B}{(\mu_A\!+\!\mu_B)^2}$ & $\frac{2}{\mu_A\!+\!\mu_B}$ & $\frac{2}{\mu_A\!+\!\mu_B}$ & $\frac{6}{(\mu_A\!+\!\mu_B)^2}$&
   8 & $B_1\!\to\! A_1$ & $\frac{\mu_A^2}{(\mu_A\!+\!\mu_B)^2}$    & $\frac{1}{\mu_A\!+\!\mu_B}$ & $\frac{2}{\mu_A\!+\!\mu_B}$ & $\frac{6}{(\mu_A\!+\!\mu_B)^2}$\\
 \hline
   4 & $A_1\!\to\! A_1$ & $\frac{\mu_A}{\mu_A\!+\!\mu_B}$      & $\frac{1}{\mu_A\!+\!\mu_B}$ & $\frac{1}{\mu_A\!+\!\mu_B}$ & $\frac{2}{(\mu_A\!+\!\mu_B)^2}$  &
   9 & $B_1\!\to\! B_0$ & $\frac{\mu_A\mu_B}{(\mu_A\!+\!\mu_B)^2}$ & $\frac{2}{\mu_A\!+\!\mu_B}$ & $\frac{2}{\mu_A\!+\!\mu_B}$ & $\frac{6}{(\mu_A\!+\!\mu_B)^2}$\\
  \hline
   5 & $A_1\!\to\! B_1$ & $\frac{\mu_B^2}{(\mu_A\!+\!\mu_B)^2}$    & $\frac{1}{\mu_A\!+\!\mu_B}$ & $\frac{2}{\mu_A\!+\!\mu_B}$ & $\frac{6}{(\mu_A\!+\!\mu_B)^2}$ &
   10 & $B_1\!\to\! B_1$ & $\frac{\mu_B}{\mu_A\!+\!\mu_B}$         & $\frac{1}{\mu_A\!+\!\mu_B}$ & $\frac{1}{\mu_A\!+\!\mu_B}$ & $\frac{2}{(\mu_A\!+\!\mu_B)^2}$\\
  \Xhline{1.2pt}
  \end{tabular}
  }
  \vspace{-1.5em}
\end{table}

\subsubsection{$l=1$}
% 1) $l=1$:
 An update from sensor A arrives at the monitor first, and the system transits from state $A_0$ to state $A_1$. In this case, the system transition probability  is given by
\begin{equation}\label{P_1}
  p_1=\mathrm{Pr}\{S_A<S_B\}=\int_0^{\infty}\int_0^{b}\mu_Ae^{-\mu_{A}a}\mu_Be^{-\mu_{B}b}{\rm d}a{\rm d}b=\frac{\mu_A}{\mu_A+\mu_B}.
\end{equation}
The expectation of the inter-arrival time $Y_1$ between two successfully received  updates at the monitor node can be calculated as
\begin{align}\label{EY_1}
  \mathbb{E}[Y_1]=\mathbb{E}[S_A|S_A<S_B]=\int_0^{\infty}\int_0^{b}\frac{a\mu_Ae^{-\mu_{A}a}\mu_Be^{-\mu_{B}b}}{\mathrm{Pr}\{S_A<S_B\}}{\rm d}a{\rm d}b
  =\frac{1}{\mu_A+\mu_B}.
\end{align}
The second-order moment of $Y_1$ is
\begin{align}\label{EYY_1}
  \mathbb{E}[Y_1^2]=\mathbb{E}[S_{A}^{2}|S_A<S_B]=\int_0^{\infty}\int_0^{b}\frac{a^2\mu_Ae^{-\mu_{A}a}\mu_Be^{-\mu_{B}b}}{\mathrm{Pr}\{S_A<S_B\}}{\rm d}a{\rm d}b
  =\frac{2}{(\mu_A+\mu_B)^2}.
\end{align}
The system transits from state $A_0$ to  state $A_1$  means that the monitor has received two consecutive updates from sensor A. In this case, the service time $T_1$ of the newly received update  is equal to the inter-arrival time  $Y_1$. Thus
\begin{equation}\label{ET_1}
  \mathbb{E}[T_1]=\mathbb{E}[Y_1]=\frac{1}{\mu_A+\mu_B}.
\end{equation}
The relevant statistics of state transition path $l=7$, i.e., the system state moves from $B_0$ to $B_1$,  can be derived similarly.

\subsubsection{$l=2$}
An update from sensor B arrives at the monitor first, and the system transits from state $A_0$ to state $B_0$. Similar to the derivation in transition path $l=1$, we can derive the relevant quantities in this context as  $p_2=\mathrm{Pr}\{S_B<S_A\}=\mu_B/(\mu_A+\mu_B)$, $\mathbb{E}[Y_2]=1/(\mu_A+\mu_B)$, and $\mathbb{E}[Y_2^2]=2/(\mu_A+\mu_B)^2$.

\begin{figure}
  \centering
\subfigure[transition path: $B_0\to A_0\to B_0$]{\label{sub_B0_A0_B0}\includegraphics[scale=0.65]{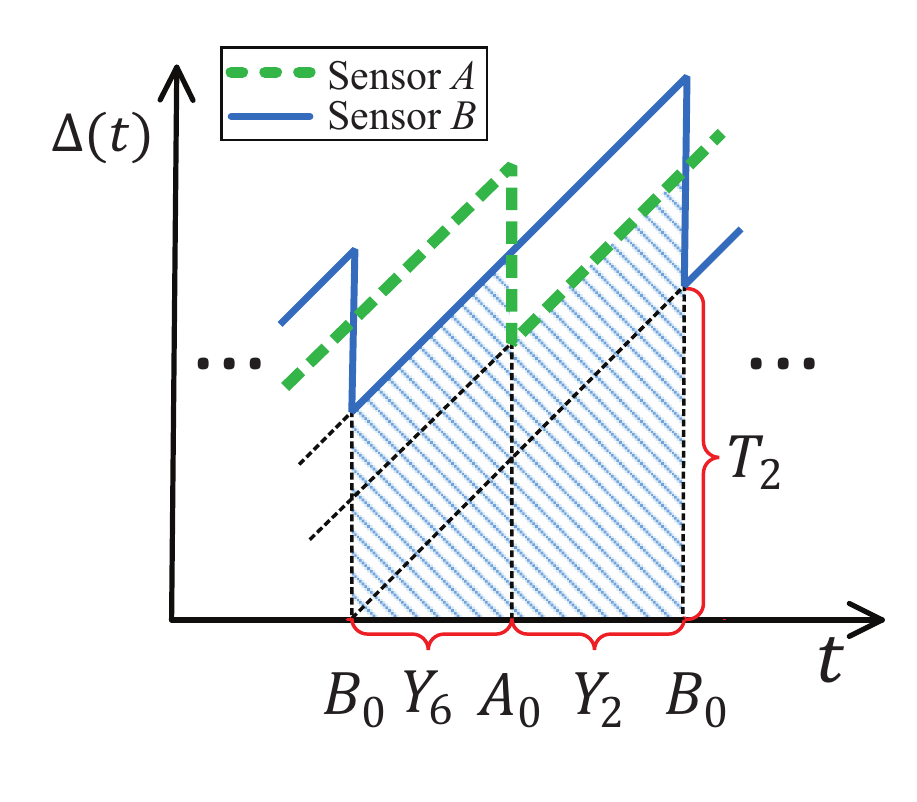}}\hspace{0.2in}
\subfigure[transition path: $A_1\to A_0\to B_0$]{\label{sub_A1_A0_B0}\includegraphics[scale=0.65]{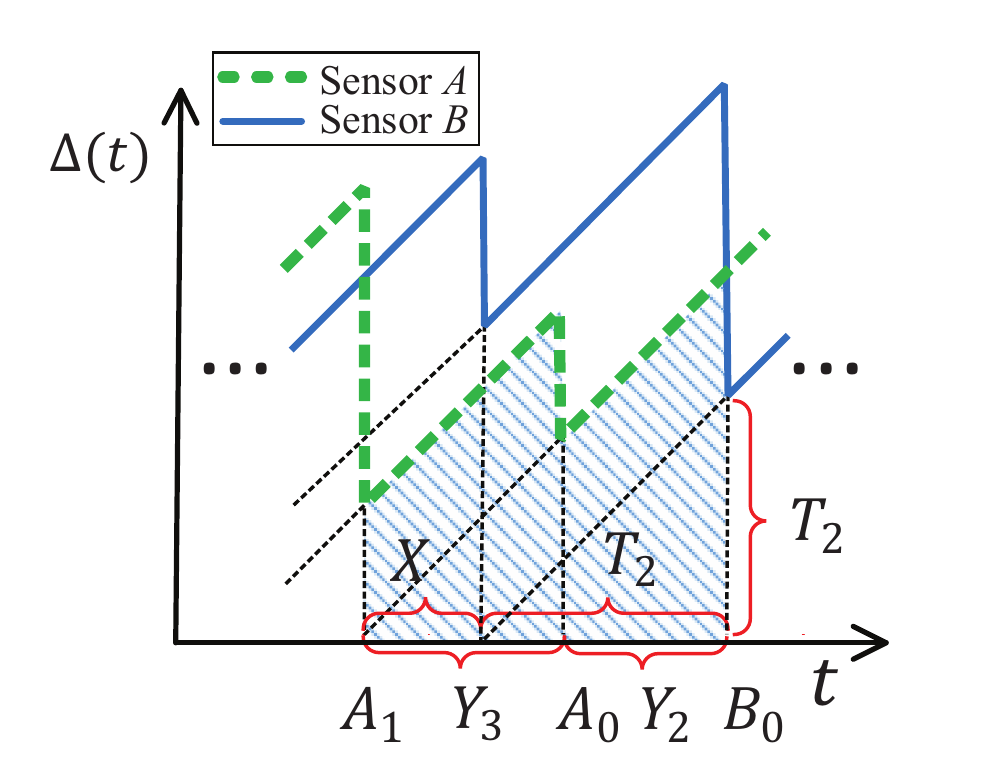}}
  \caption{An evolution example of the instantaneous AoI when the system state transition path is $A_0\to B_0$.}
  \label{MD_B0_A0_B0}
 \vspace{-1.5em}
\end{figure}

It is worth noting that the service time $T_2$ of the newly received update should be divided into two cases for discussion, i.e., $B_0\to A_0\to B_0$ and $A_1\to A_0\to B_0$, as shown in Fig.~\ref{MD_B0_A0_B0}.
For the system state transition path $B_0\to A_0\to B_0$, we can see from Fig.~\ref{sub_B0_A0_B0} that the service time $T_2$ of the newly received update is equal to the sum of the inter-arrival  time $Y_6$ of transition path $l = 6$ and the inter-arrival  time $Y_2$ of transition path $l = 2$.
Hence, the average service time can be further expressed as \footnote{Similar to transition path $l=1$, one can get that $p_6=\mathrm{Pr}\{S_A<S_B\}=\mu_A/(\mu_A+\mu_B)$ and $\mathbb{E}[Y_6]=1/(\mu_A+\mu_B)$.}
 \begin{equation}
   \mathbb{E}[T_2]=\mathbb{E}[Y_6+Y_2]=\mathbb{E}[Y_6]+\mathbb{E}[Y_2]=\frac{2}{\mu_A+\mu_B}.
 \end{equation}

For the system state transition path $A_1\to A_0\to B_0$, we can see from Fig.~\ref{sub_A1_A0_B0} that the service time $T_2$ of the newly received update consists of the inter-arrival  time $Y_2$ of transition path $l = 2$ and a part of the  inter-arrival  time $Y_3$ of  transition path $l = 3$.
The conditional probability density function of the random variable $X$ in Fig.~\ref{sub_A1_A0_B0} is given by
\begin{align}\label{CPDF}
  f_{X|Y_{3},Y_{2}}\left(x|y_3,y_2\right)
  =\frac{\mu_Be^{-\mu_Bx}\mu_Be^{-\mu_B(y_3+y_2-x)}}{\int_0^{y_3}\mu_Be^{-\mu_Bx}\mu_Be^{-\mu_B(y_3+y_2-x)}{\rm d}x}
  =\frac{1}{y_3}.
\end{align}
Hence, the conditional expectation of $X$ takes the following form
\begin{equation}\label{EX}
  \mathbb{E}[X|Y_3=y_3,Y_2=y_2]=\int_0^{y_3}xf_{X|Y_{3},Y_{2}}\left(x|y_3,y_2\right){\rm d}x=\frac{y_3}{2}.
\end{equation}
Deconditioning the above yields
 \begin{equation}
\mathbb{E}[X]=\mathbb{E}\left[\mathbb{E}[X|Y_3,Y_2]\right]=\int_0^{\infty}\int_0^{\infty}\frac{y_3}{2}f_{Y_3,Y_2}(y_3,y_2){\rm d}y_3{\rm d}y_2
=\frac{1}{2}\mathbb{E}[Y_3].
 \end{equation}
Thus, the average service time is given as \footnote{$\mathbb{E}[Y_3]$ is given in  transition $l = 3$.}
 \begin{align}
   \mathbb{E}[T_2]=\mathbb{E}[Y_3+Y_2-X]
   =\mathbb{E}[Y_3]+\mathbb{E}[Y_2]-\mathbb{E}[X]
   =\frac{2}{\mu_A+\mu_B}.
 \end{align}

In summary, when the system transits from state $A_0$ to $B_0$, the average service time of the newly received updates is $\mathbb{E}[T_2]=2/(\mu_A+\mu_B)$.
The statistics of transition path $l=6$  can be derived in a similar way.

\subsubsection{$l=3$}
An update from sensor B arrives at the monitor but the AoI of this update is higher than the monitor's instantaneous AoI.  Hence, the arrival of this update can not reduce the AoI observed at the monitor.
Before sensor B successfully transmits the next update, sensor A successfully sends an update to the monitor.
Then, the AoI at the monitor can be refreshed and the system transits from state $A_1$ to state $A_0$ as shown in Fig.~\ref{sub_A1_A0_B0}.

Let us define $S_{BB}$ as the sum of service time of two consecutive updates from sensor B. Then, $S_{BB}$ obeys the second-order Erlang distribution with parameter $\mu_B$, where the probability density function of $S_{BB}$ is $f_{S_{BB}}(x)=\mu_B^2xe^{-\mu_Bx}$ with $x\geq0$.
Similarly, we can construct a random variable $S_{AA}$ as the sum of two consecutive updates from sensor A, which obeys second-order Erlang distribution with parameter $\mu_A$. Consequently, the system transition probability can be computed as
\begin{align}\label{P_3}
  p_3\!=\!\mathrm{Pr}\{S_B\!<\!S_A\!<\!S_{BB}\}
  \!=\!\int_0^{\infty}\!\int_0^{\infty}\!\int_b^{x}\mu_Ae^{-\mu_{A}a}\mu_Be^{-\mu_{B}b}\mu_B^{2}xe^{-\mu_{B}x}{\rm d}a{\rm d}b{\rm d}x
 \!=\!\frac{\mu_A\mu_B}{(\mu_A\!+\!\mu_B)^2}.
 \end{align}
The first-order and second-order moments of the inter-arrival time $Y_3$ can be respectively calculated by the following
\begin{align}\label{EY_3}
  \mathbb{E}[Y_3]&=\mathbb{E}[S_A|S_B<S_A<S_{BB}]
  =\int_0^{\infty}\int_0^{\infty}\int_b^{x}\frac{a\mu_Ae^{-\mu_{A}a}\mu_Be^{-\mu_{B}b}\mu_B^{2}xe^{-\mu_{B}x}}{\mathrm{Pr}\{S_B<S_A<S_{BB}\}}{\rm d}a{\rm d}b{\rm d}x\nonumber\\
  &=\frac{2}{\mu_A+\mu_B},\\
  \mathbb{E}[Y_3^2]&=\mathbb{E}[S_A^2|S_B<S_A<S_{BB}]=\frac{6}{(\mu_A+\mu_B)^2}.
\end{align}
The transition of system from  state $A_1$ to  state $A_0$  means that the monitor successfully receives two consecutive updates from sensor A. Hence, the service time of the newly received update is equal to the interval-arrival time, i.e.,
\begin{equation}\label{ET_3}
  \mathbb{E}[T_3]=\mathbb{E}[Y_3]=\frac{2}{\mu_A+\mu_B}.
\end{equation}
The relevant statistics of transition path $l = 9$ can be derived in a similar way.

\subsubsection{$l=4$}
 When the system is in state $A_1$, if an update from sensor A arrives at the monitor first, the system remains in state $A_1$. Similar to the case in  transition $l=1$, we have $p_4=\mu_A/(\mu_A+\mu_B)$, $\mathbb{E}[Y_4]=1/(\mu_A+\mu_B)$, $\mathbb{E}[Y_4^2]=2/(\mu_A+\mu_B)$, and $\mathbb{E}[T_4]=1/(\mu_A+\mu_B)$.
 The relevant statistics of transition path $l = 10$ can be derived similarly.
\subsubsection{$l=5$}
\begin{figure}
  \centering
  \includegraphics[scale=0.65]{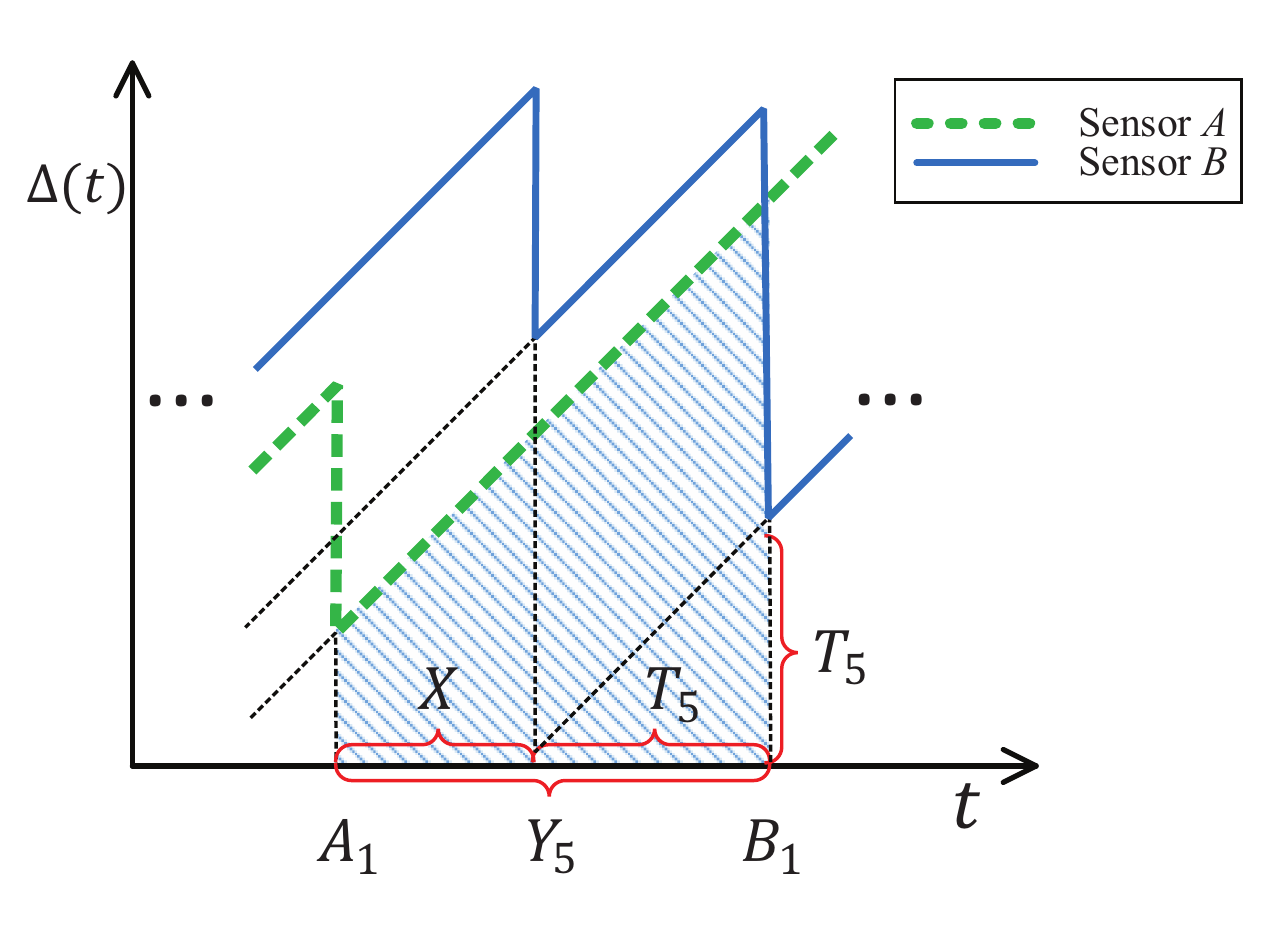}
  \caption{An evolution example of the instantaneous AoI when the system state transition path is $A_1\to B_1$.}
  \label{A1_B1}
  \vspace{-1.5em}
\end{figure}

 Two consecutive updates from sensor B arrive at the monitor, and the system transits from state $A_1$ to state $B_1$.
Fig.~\ref{A1_B1} shows an evolution example of the instantaneous AoI when the system transition path follows $A_1\to B_1$.
In this case, the system transition probability is given by
\begin{align}\label{P_5}
  p_5=\mathrm{Pr}\{S_{BB}<S_A\}=\int_0^{\infty}\int_0^{a}\mu_B^{2}xe^{-\mu_{B}x}\mu_Ae^{-\mu_{A}a}{\rm d}x{\rm d}a
  =\frac{\mu_B^2}{(\mu_A+\mu_B)^2}.
\end{align}
The first and second moments of the inter-arrival time $Y_5$ can be respectively
expressed as
\begin{align}\label{EY_5}
  \mathbb{E}[Y_5]&=\mathbb{E}[S_{BB}|S_{BB}<S_A]=\int_0^{\infty}\int_0^{a}\frac{x\mu_B^{2}xe^{-\mu_{B}x}\mu_Ae^{-\mu_{A}a}}{\mathrm{Pr}\{S_{BB}<S_A\}}{\rm d}x{\rm d}a
  =\frac{2}{\mu_A+\mu_B},\\
  \mathbb{E}[Y_5^2]&=\mathbb{E}[S_{BB}^2|S_{BB}<S_A]=\frac{6}{(\mu_A+\mu_B)^2}.
\end{align}
%\begin{align}\label{EYY_5}
%  \mathbb{E}[Y_5^2]=\mathbb{E}[S_{BB}^2|S_{BB}<S_A]
%  =\int_0^{\infty}\int_0^{a}\frac{x^2\mu_B^{2}xe^{-\mu_{B}x}\mu_Ae^{-\mu_{A}a}}{\mathrm{Pr}\{S_{BB}<S_A\}}{\rm d}x{\rm d}a
%  =\frac{6}{(\mu_A+\mu_B)^2}.
%\end{align}

As we can see from Fig.~\ref{A1_B1}, the service time $T_5$ of the newly received update is  part of the inter-arrival time $Y_5$.
Besides, the conditional probability density function of the random variable $X$ in Fig.~\ref{A1_B1} can be calculated as
\begin{align}\label{B1_A1_CPDF}
  f_{X|Y_5}\left(x|y_5\right)=\frac{\mu_Ae^{-\mu_Ax}\mu_Ae^{-\mu_A(y-x)}}{\int_0^{y_5}\mu_Ae^{-\mu_Ax}\mu_Ae^{-\mu_A(y-x)}{\rm d}x}
  =\frac{1}{y_5}.
\end{align}
We can compute the conditional expectation of $X$ as
\begin{equation}
 \mathbb{E}[X|Y_5=y_5]\!=\!\int_{0}^{y_5}xf_{X|Y_5=y_5}\left(x|Y_5=y_5\right){\rm d}x=\frac{y_5}{2}.
\end{equation}
And deconditioning the above yields
\begin{equation}
  \mathbb{E}[X]=\mathbb{E}\left[\mathbb{E}[X|Y_5]\right]=\frac{1}{2}\mathbb{E}[Y_5].
\end{equation}
Therefore, the mean service time is
\begin{equation}
  \mathbb{E}[T_5]=\mathbb{E}[Y_5-X]=\mathbb{E}[Y_5]-\mathbb{E}[X]=\frac{1}{\mu_A+\mu_B}.
\end{equation}
The relevant statistics of transition path $l=8$  can be derived similarly.

We index the states $A_0$, $A_1$, $B_0$, and $B_1$ as states $1$, 2, 3, and 4, respectively, and express the state transition matrix as
\begin{equation}       %开始数学环境
\mathbf{\Lambda}=\left[                 %左括号
  \begin{array}{cccc}   %该矩阵一共3列，每一列都居中放置
0 & \frac{\mu_A}{\mu_A+\mu_B} & \frac{\mu_B}{\mu_A+\mu_B} & 0 \\ \vspace{1ex} % 控制行距
\frac{\mu_A\mu_B}{(\mu_A+\mu_B)^2} & \frac{\mu_A}{\mu_A+\mu_B} & 0 & \frac{(\mu_B)^2}{(\mu_A+\mu_B)^2} \\ \vspace{1ex}
\frac{\mu_A}{\mu_A+\mu_B} & 0 & 0 & \frac{\mu_B}{\mu_A+\mu_B} \\ \vspace{1ex}
0 & \frac{(\mu_A)^2}{(\mu_A+\mu_B)^2} & \frac{\mu_A\mu_B}{(\mu_A+\mu_B)^2} &  \frac{\mu_B}{\mu_A+\mu_B}\\
  \end{array}
\right],                 %右括号
\end{equation}
where the $(i,j)\text{th}$ entry represents the transition probability of the system transitioning from state $i$ to $j$.
Let $\boldsymbol{\pi}=(\pi_1,\pi_2,\pi_3,\pi_4)$ denote a vector of steady-state probabilities of the system. Then, we have
\begin{equation}\label{steady state F}
  \boldsymbol{\pi}=\boldsymbol{\pi}\mathbf{\Lambda}, \ \rm and \ \sum_{j=1}^4\pi_j=1.
\end{equation}
This fixed-point system of equations can be solved by the following
\begin{align}\label{steady state}
  \pi_1=\frac{\mu_A^2\mu_B}{\xi(\mu_A+\mu_B)},\quad
  \pi_2=\frac{\mu_A^2}{\xi},\quad
  \pi_3=\frac{\mu_A\mu_B^2}{\xi(\mu_A+\mu_B)},\quad
  \pi_4=\frac{\mu_B^2}{\xi},
\end{align}
where $\xi=\mu_A^2+\mu_A\mu_B+\mu_B^2$.
Combining the transition probability in Table \ref{State_pro} and the steady-state probability in (\ref{steady state}),  the occurrence probability $P_l$ for transition $l$ in the system can be written as $P_1\!=\!\frac{\mu_A^3\mu_B}{\xi(\mu_A\!+\!\mu_B)^2}$, $P_2\!=\!\frac{\mu_A^2\mu_B^2}{\xi(\mu_A\!+\!\mu_B)^2}$, $P_3\!=\!\frac{\mu_A^3\mu_B}{\xi(\mu_A\!+\!\mu_B)^2}$, $P_4\!=\!\frac{\mu_A^3}{\xi(\mu_A\!+\!\mu_B)}$, $P_5\!=\!\frac{\mu_A^2\mu_B^2}{\xi(\mu_A\!+\!\mu_B)^2}$, $P_6\!=\!\frac{\mu_A^2\mu_B^2}{\xi(\mu_A\!+\!\mu_B)^2}$, $P_7\!=\!\frac{\mu_A\mu_B^3}{\xi(\mu_A\!+\!\mu_B)^2}$, $P_8\!=\!\frac{\mu_A^2\mu_B^2}{\xi(\mu_A\!+\!\mu_B)^2}$, $P_9\!=\!\frac{\mu_A\mu_B^3}{\xi(\mu_A\!+\!\mu_B)^2}$, and $P_{10}\!=\!\frac{\mu_B^3}{\xi(\mu_A\!+\!\mu_B)}$, where $\sum_{l=1}^{10}P_l=1$.

As such, the average inter-arrival time $\mathbb{E}[Y]$ is given by
\begin{equation}\label{EY}
  \mathbb{E}[Y]=\sum_{l\in\mathcal{L}}P_l\mathbb{E}[Y_l]=\frac{\mu_A+\mu_B}{\mu_A^2+\mu_A\mu_B+\mu_B^2}.
\end{equation}
Likewise, we can obtain the average service time
\begin{equation}\label{ET}
  \mathbb{E}[T]=\sum_{l\in\mathcal{L}}P_l\mathbb{E}[T_l]=\frac{\mu_A+\mu_B}{\mu_A^2+\mu_A\mu_B+\mu_B^2}.
\end{equation}
From Fig.~\ref{MM_AoI_evolution}, we note that the PAoI observed at the moment of successfully  receiving the $n\text{th}$ update is $A_{(n)}=T_{(n-1)}+Y_{(n)}$.
According to the independence of $T_{(n-1)}$ and $Y_{(n)}$, the average PAoI can be expressed as
\begin{equation}\label{MM_paoi_shunshi}
\Delta_{\text{M-M}}^{\text{peak}}=\mathbb{E}[T]+\mathbb{E}[Y].
\end{equation}
Substituting (\ref{EY}) and (\ref{ET}) into (\ref{MM_paoi_shunshi}) leads to Proposition \ref{prop_mm_paoi}.

\subsection{Average AoI Analysis}\label{proof_MM_aoi}
Let us denote the number of successfully received updates in the time interva $(0, \tau)$ as $N_\tau$.
Then,  the time average AoI $\Delta$ can be expressed as
\begin{equation}\label{time_aver_AoI}
  \Delta=\lim_{\tau\to\infty}\frac{1}{\tau}\sum_{n=1}^{N_\tau}Q_{(n)}=\lim_{\tau\to\infty}\frac{N_\tau}{\tau}\frac{1}{N_\tau}\sum_{n=1}^{N_\tau}Q_{(n)}
  =\frac{\mathbb{E}[Q_{(n)}]}{\mathbb{E}[Y]}.
\end{equation}
From Fig.~\ref{MM_AoI_evolution} we can derive the following
\begin{align}\label{EQn}
  \mathbb{E}[Q_{(n)}]=\mathbb{E}\left[\frac{\left(T_{(n-1)}+Y_{(n)}+T_{(n-1)}\right)Y_{(n)}}{2}\right]
  =\mathbb{E}[T_{(n-1)}Y_{(n)}]+\frac{1}{2}\mathbb{E}[Y_{(n)}^2].
\end{align}
Therefore, the area $Q_{(n)}$  of the trapezoid associated with the $n\text{th}$ successfully received update not only depends on the inter-arrival time $Y_{(n)}$ of the current update, but also the service time $T_{(n-1)}$ of the previous successfully received update, i.e., $Q_{(n)}$ is determined by the previous transition path and the current transition path.
As can be seen from Fig.~\ref{MM_state_transition}, the M-M system has 10 transition paths. According to the difference of two consecutive transfer paths, we can divide the system transition state into 26 cases.
Table \ref{ETEY} summarizes the statistics related to 13 of these system transition paths.
In Table \ref{ETEY}, $P_c$ represents the probability of transition  $c$: $q_c\to q_c^{\prime}\to q_c^{\prime\prime}$ in the system.
 $\mathbb{E}(TY_c)$ is the mean  of the product of the service time of state transition $q_c\to q_c^{\prime}$ and the inter-arrival time of transition $q_c^{\prime}\to q_c^{\prime\prime}$,
and $\mathbb{E}(Y_c^2)$ is the second-order moment of the inter-arrival time of transition $q_c^{\prime}\to q_c^{\prime\prime}$. The statistics in Table \ref{ETEY} can be derived from Table \ref{State_pro} and the  probability $P_l$ of transition path $l$.
Similarly, the statistics related to the remaining 13 state transitions can be obtained.
The details are omitted due to the space limitation.
\begin{table}
  \centering
  \caption{Statistics corresponding to the state transitions for the M-M system.}\label{ETEY}
 \renewcommand\arraystretch{1.5}
 \resizebox{\textwidth}{!}{
  \begin{tabular}{!{\vrule width1.2pt}c|c|c|c|c||c|c|c|c|c!{\vrule width1.2pt}}
  \Xhline{1.2pt}
  $c$ & $q_c\!\to\!q_c^{\prime}\!\to\!q_c^{\prime\prime}$  & $P_c$   & $\mathbb{E}(TY_c)$     & $\mathbb{E}(Y_c^2)$ &  7 & $A_1\!\to\!A_0\!\to\!B_0$ & $\frac{\mu_A^3\mu_B^2}{\xi(\mu_A+\mu_B)^3}$     & $\frac{2}{(\mu_A+\mu_B)^2}$ & $\frac{2}{(\mu_A+\mu_B)^2}$ \\
  \hline
    1 & $A_0\!\to\!A_1\!\to\!A_0$ & $\frac{\mu_A^4\mu_B^2}{\xi(\mu_A+\mu_B)^4}$   & $\frac{2}{(\mu_A+\mu_B)^2}$ & $\frac{6}{(\mu_A+\mu_B)^2}$     &
    8 & $A_1\!\to\!A_1\!\to\!A_0$ & $\frac{\mu_A^4\mu_B}{\xi(\mu_A+\mu_B)^3}$       & $\frac{2}{(\mu_A+\mu_B)^2}$ & $\frac{6}{(\mu_A+\mu_B)^2}$   \\
  \hline
  2 & $A_0\!\to\!A_1\!\to\!A_1$ & $\frac{\mu_A^4\mu_B}{\xi(\mu_A+\mu_B)^3}$       & $\frac{1}{(\mu_A+\mu_B)^2}$ & $\frac{2}{(\mu_A+\mu_B)^2}$     &
   9 & $A_1\!\to\!A_1\!\to\!A_1$ & $\frac{\mu_A^4}{\xi(\mu_A+\mu_B)^2}$            & $\frac{1}{(\mu_A+\mu_B)^2}$ & $\frac{2}{(\mu_A+\mu_B)^2}$    \\
  \hline
  3 & $A_0\!\to\!A_1\!\to\!B_1$ & $\frac{\mu_A^3\mu_B^3}{\xi(\mu_A+\mu_B)^4}$     & $\frac{2}{(\mu_A+\mu_B)^2}$ & $\frac{6}{(\mu_A+\mu_B)^2}$     &
  10 & $A_1\!\to\!A_1\!\to\!B_1$ & $\frac{\mu_A^3\mu_B^2}{\xi(\mu_A+\mu_B)^3}$    & $\frac{2}{(\mu_A+\mu_B)^2}$ & $\frac{6}{(\mu_A+\mu_B)^2}$     \\
  \hline
  4 & $A_0\!\to\!B_0\!\to\!A_0$ & $\frac{\mu_A^3\mu_B^2}{\xi(\mu_A+\mu_B)^3}$     & $\frac{2}{(\mu_A+\mu_B)^2}$ & $\frac{2}{(\mu_A+\mu_B)^2}$     &
  11 & $A_1\!\to\!B_1\!\to\!A_1$ & $\frac{\mu_A^4\mu_B^2}{\xi(\mu_A+\mu_B)^4}$    & $\frac{2}{(\mu_A+\mu_B)^2}$ & $\frac{6}{(\mu_A+\mu_B)^2}$     \\
  \hline
  5 & $A_0\!\to\!B_0\!\to\!B_1$ & $\frac{\mu_A^2\mu_B^3}{\xi(\mu_A+\mu_B)^3}$     & $\frac{2}{(\mu_A+\mu_B)^2}$ & $\frac{2}{(\mu_A+\mu_B)^2}$     &
  12 & $A_1\!\to\!B_1\!\to\!B_0$ & $\frac{\mu_A^3\mu_B^3}{\xi(\mu_A+\mu_B)^4}$    & $\frac{2}{(\mu_A+\mu_B)^2}$ & $\frac{6}{(\mu_A+\mu_B)^2}$    \\
  \hline
  6 & $A_1\!\to\!A_0\!\to\!A_1$ & $\frac{\mu_A^4\mu_B}{\xi(\mu_A+\mu_B)^3}$       & $\frac{2}{(\mu_A+\mu_B)^2}$ & $\frac{2}{(\mu_A+\mu_B)^2}$     &
  13 & $A_1\!\to\!B_1\!\to\!B_1$ & $\frac{\mu_A^2\mu_B^3}{\xi(\mu_A+\mu_B)^3}$    & $\frac{1}{(\mu_A+\mu_B)^2}$ & $\frac{2}{(\mu_A+\mu_B)^2}$    \\
  \Xhline{1.2pt}
  \end{tabular}
  }
  \vspace{-1.5em}
\end{table}

Note that the average AoI of the M-M system can be obtained by the following
\begin{equation}\label{MM_AoI_derived}
  \Delta_{\text{M-M}}=\frac{\sum_{c=1}^{26}P_{c}\left(\mathbb{E}[TY_c]+\frac{1}{2}\mathbb{E}[Y_c^2]\right)}{\mathbb{E}[Y]}.
\end{equation}
Combining Table \ref{ETEY}, (\ref{EY}), and (\ref{MM_AoI_derived}), we arrive at the analytical expression of average AoI of the M-M system in  Proposition \ref{prop_mm_aoi}.

\section{AoI of the M-D System}\label{sec:MD}
\allowdisplaybreaks
In this section, we analyze the average PAoI and AoI of the M-D system, in which the service time $T$ of sensor B is deterministic and the service time of sensor A follows the exponential distribution with service rate $\mu$.
\subsection{System State Analysis}
\allowdisplaybreaks
Since the service time of sensor B in the M-D system is deterministic, we can evaluate the average AoI of the M-D system by analyzing the sawtooth AoI waveform during a service period $T$ of sensor B.
Let $\widetilde{N}(T)$ and $N(T)$ denote the number of updates sent by sensor A in the previous period and the current period, respectively.
Because the service time of sensor A is exponentially distributed, $\widetilde{N}(T)$ and $N(T)$ are Poisson processes.
As we can see from Fig.~\ref{MM_AoI_evolution}, for a service period of sensor B, the area covered by the sawtooth AoI waveform  is determined by the number of updates sent by sensor A in the current period and the previous period.
We use the notation $(k,n)$ to represent the state of M-D system, where $k\in \mathbb{N}$ and $n\in \mathbb{N}$ represent the number of updates transmitted by sensor A in the previous period and the current period, respectively.
%As we can see from Fig.~\ref{MD_evolution}, during the period service time $T$ of sensor B, sensor A may have successfully sent several updates, or it may not have sent any updates.
The probability  of the M-D system in state $(k,n)$ can be expressed as
\begin{equation}\label{possion}
  \mathrm{Pr}\{\widetilde{N}(T)=k,N(T)=n\}=\frac{(\mu T)^k}{k!}e^{-\mu T}\frac{(\mu T)^n}{n!}e^{-\mu T}.
\end{equation}

According to the number of updates transmitted by  sensor A in the current service period and  the previous period, we can categorize the system state into the following cases.

\subsubsection{State $(0,0)$}
\begin{figure}
  \centering
\subfigure[state $(0,0)$]{\label{sub_MD_00}\includegraphics[scale=0.25]{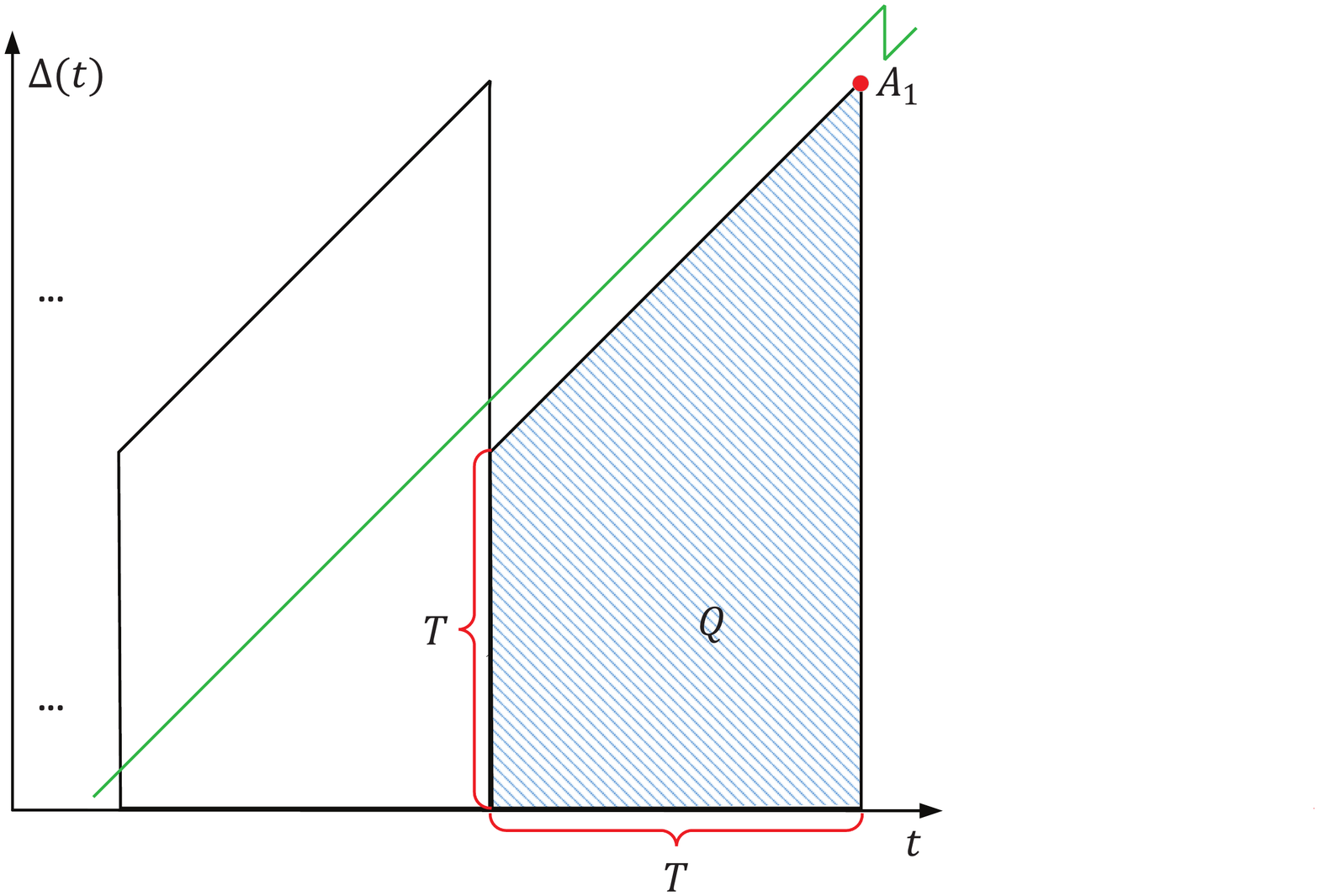}}\hspace{0.2in}
\subfigure[state $(0,1)$]{\label{sub_MD_01}\includegraphics[scale=0.25]{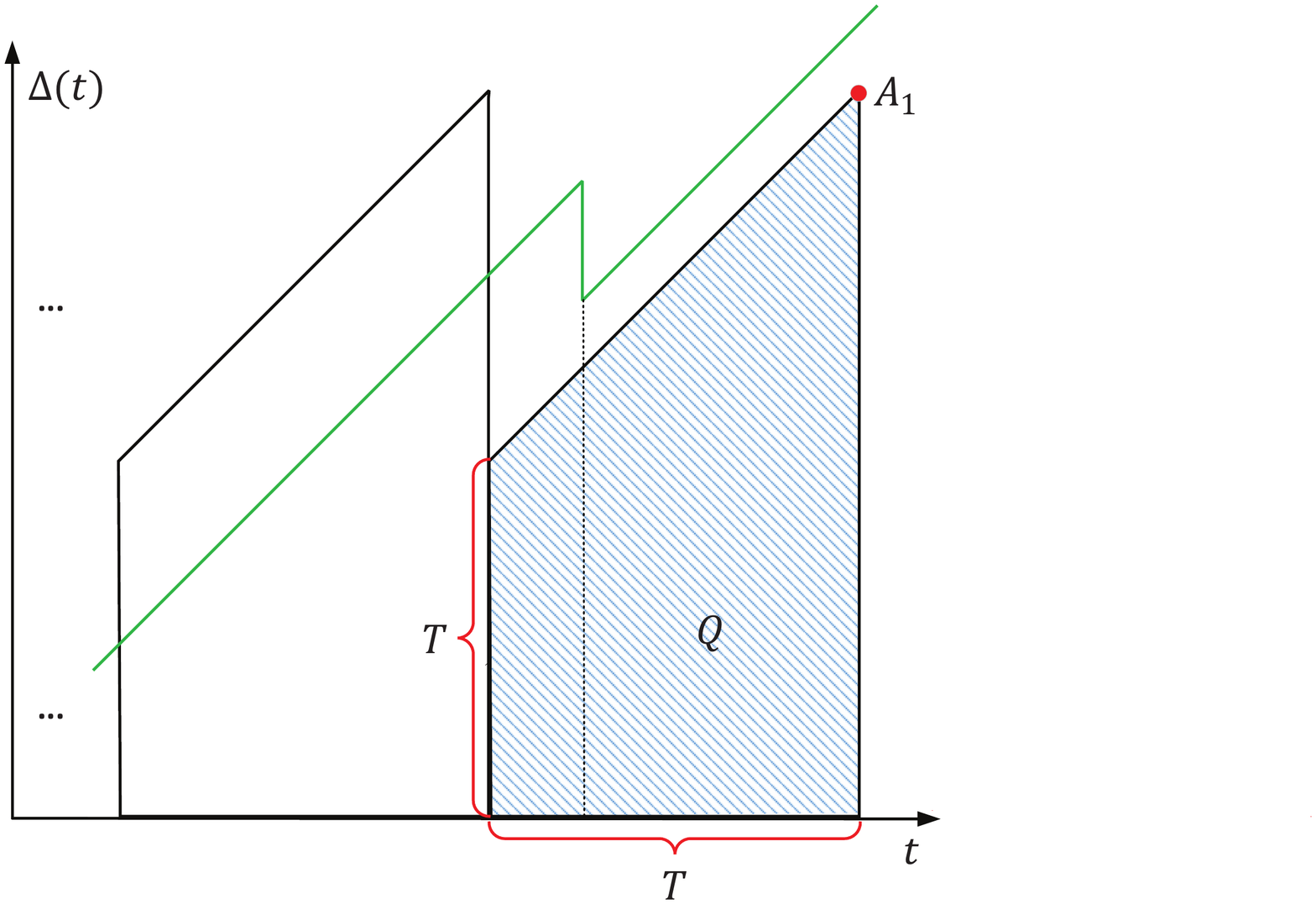}}\hspace{0.2in}
\subfigure[state $(0,n)$]{\label{sub_MD_0n}\includegraphics[scale=0.25]{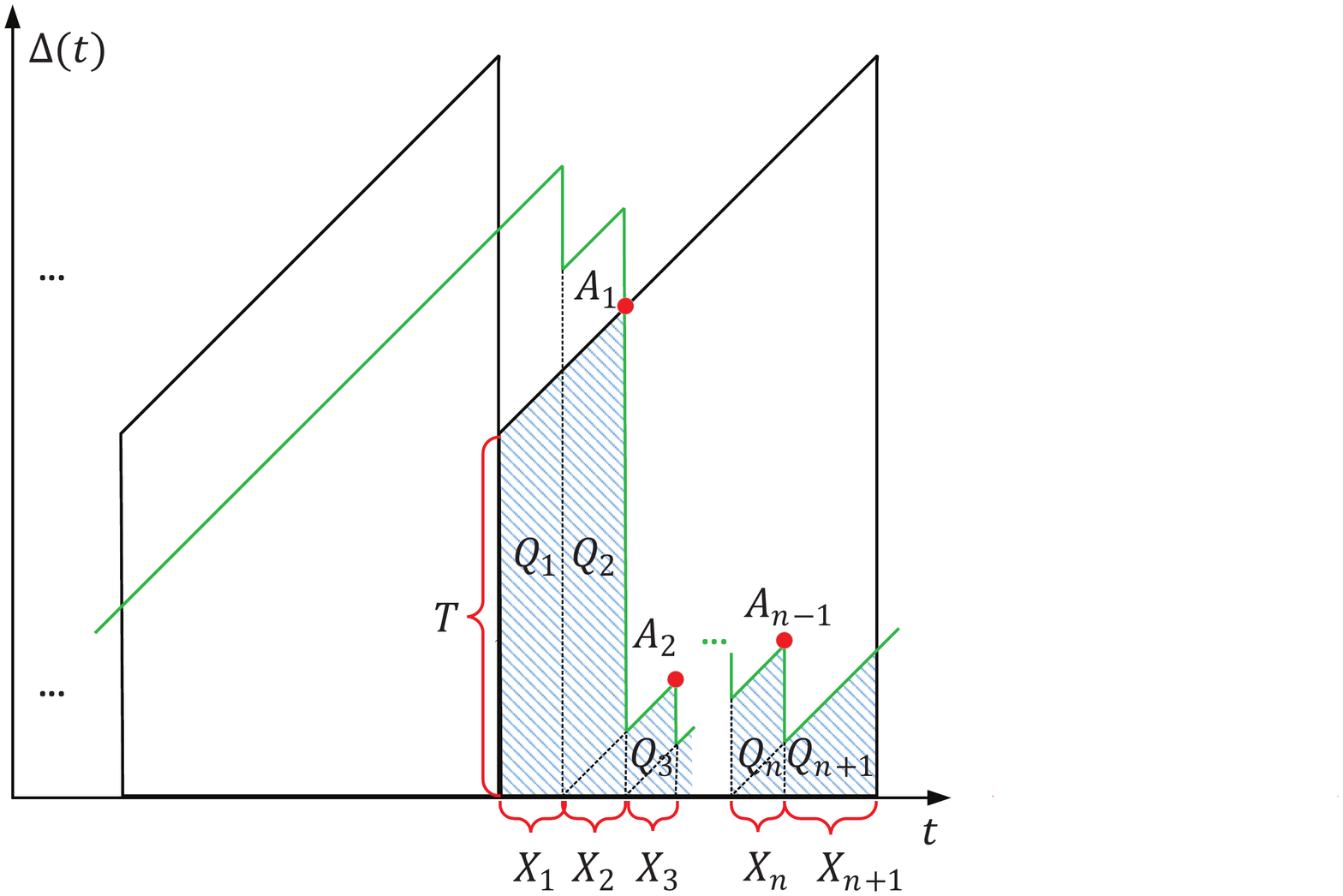}}
\subfigure[state $(k,0)$]{\label{sub_MD_k0}\includegraphics[scale=0.25]{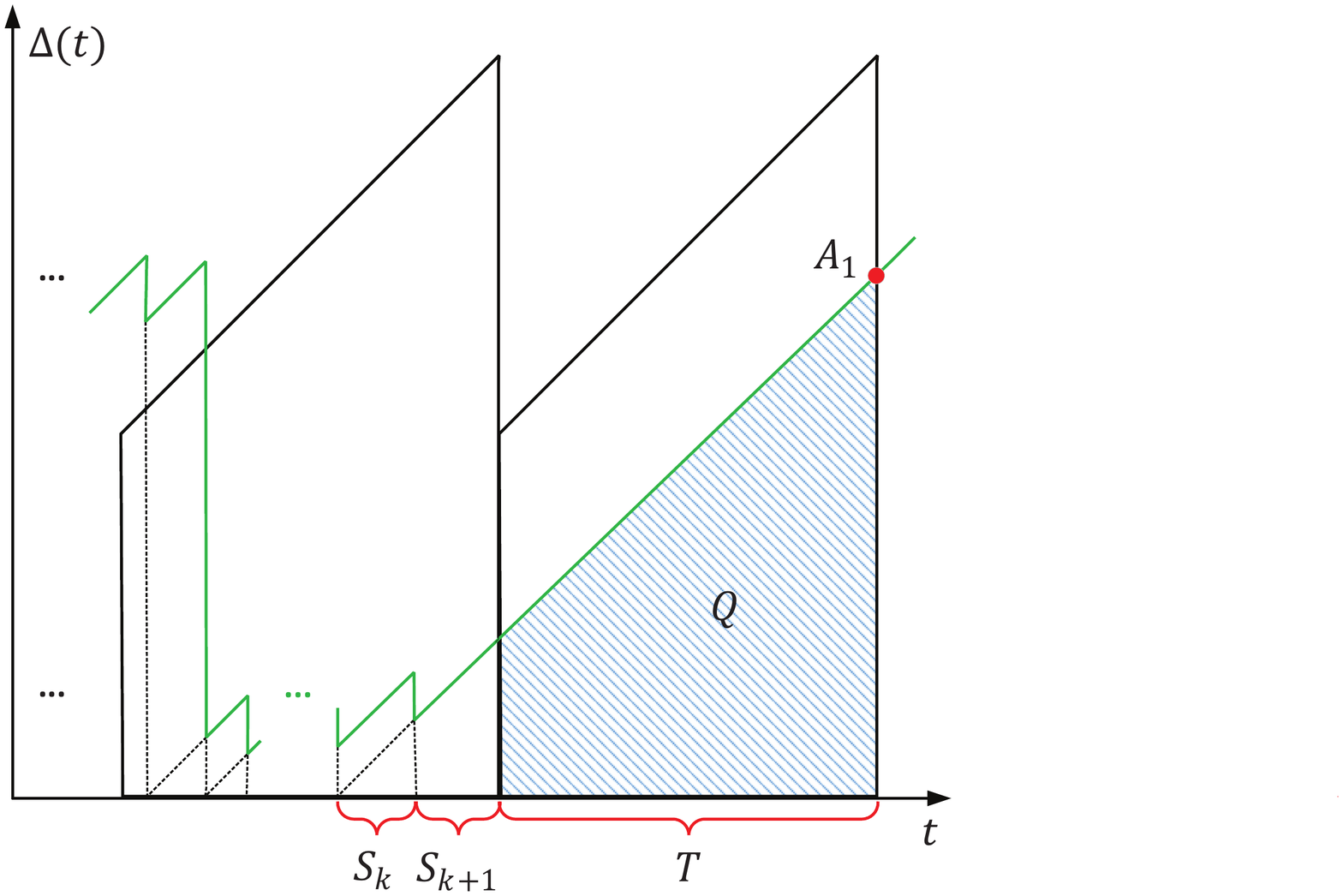}}\hspace{0.2in}
\subfigure[state $(k,1)$]{\label{sub_MD_k1}\includegraphics[scale=0.25]{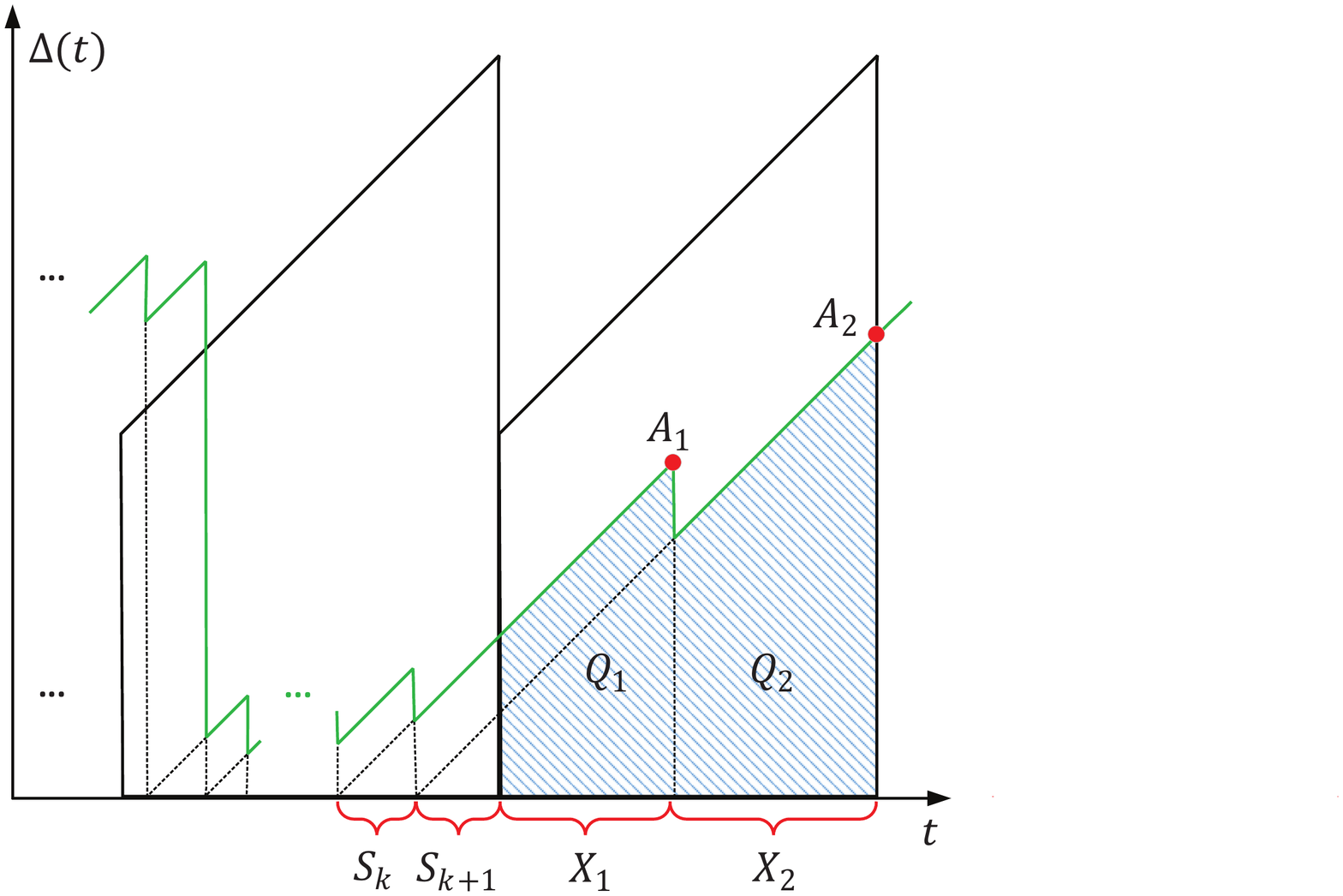}}\hspace{0.2in}
\subfigure[state $(k,n)$]{\label{sub_MD_kn}\includegraphics[scale=0.25]{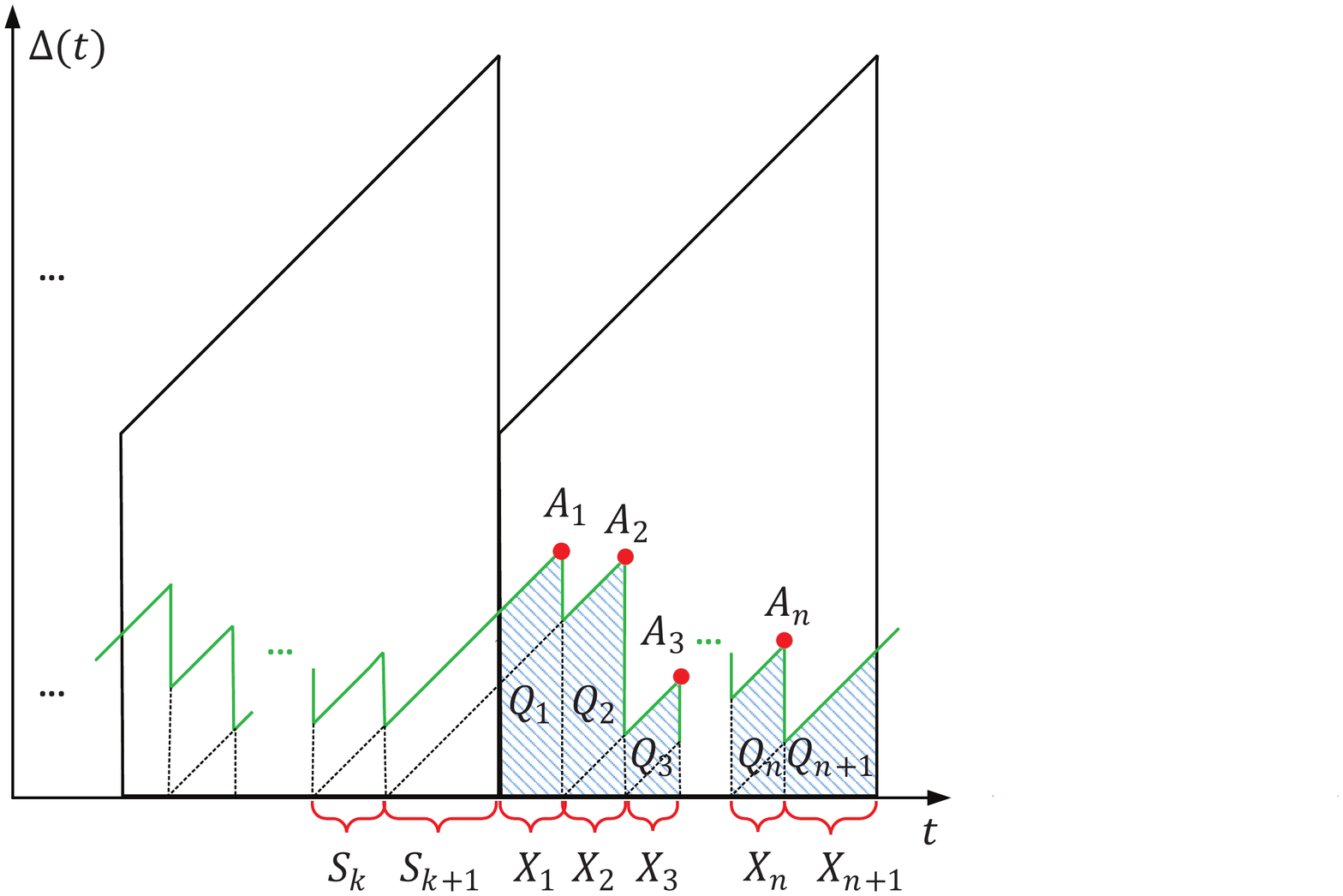}}
  \caption{Evolution examples of the instantaneous AoI for the M-D system in typical states.}
  \label{B0_A0_B0}
   \vspace{-1.5em}
\end{figure}

During the previous service period and the  current service period of sensor B,  no update from sensor A arrives at the monitor.
Fig.~\ref{sub_MD_00} shows an example of AoI in this case. Note that the monitor only received an update from sensor B during the current service period.
Therefore, the PAoI is
\begin{equation}\label{PA_00}
  A=T+T=2T.
\end{equation}

Moreover, the area covered by the AoI waveform at the monitor, i.e., the shaded part in Fig.~\ref{sub_MD_00}, is
\begin{equation}\label{Q_00}
  Q=\frac{(T+T+T)T}{2}=\frac{3T^2}{2}.
\end{equation}

\subsubsection{State $(0,1)$ }
Sensor A did not sent an update in the previous service period but transmits one update in the current period, as shown in Fig.~\ref{sub_MD_01}.
In the current service period, the monitor receives two updates in total. One from  sensor A and the other from sensor B. Due to the long duration of service time, the update of sensor A is stale when it arrives at the destination. Therefore, such an update does not refresh the AoI on the monitor.
Consequently,   the sum average of PAoI in this state is
\begin{equation}\label{01_A}
  \mathbb{E}\left[A|(0,1)\right]=T+T=2T.
\end{equation}

The average area under AoI curve of the M-D system in state $(0,1)$ is given by
\begin{equation}\label{01_Q}
  \mathbb{E}\left[Q|(0,1)\right]=\frac{T+T+T}{T}=\frac{3T^2}{2}.
\end{equation}

\subsubsection{State $(0,n)$}
Sensor A did not update in the previous service period, and transmits $n(n\ge2)$ updates in the current period, as shown in Fig.~\ref{sub_MD_0n}.
Similar to state $(0,1)$,  the first packet received from sensor A is obsolete because of the long service time.
Therefore, the arrival of the first packet can not refresh the AoI on the monitor and the monitor totally receives $n-1$ valid updates in the current service period.
Let $X_{i}(i=1,2,...,n)$ denote the service time of the $i\text{th}$ update.
Within a period $T$, the remaining time after sensor A successfully sent $n$ updates is $X_{n+1}$.
According to the properties of Poisson process, the joint probability density function of $(X_{1},...,X_{n})$ is
\begin{align}\label{0n_jpdf}
  f_{X_{1},...,X_{n}}(x_{1},...,x_{n}|N(T)=n)
  =\frac{\mu e^{-\mu x_{1}}\cdots\mu e^{-\mu x_{n}}e^{-\mu\left(T-\sum_{i=1}^{n}x_{i}\right)}}{\frac{(\mu T)^{n}}{n!}e^{-\mu T}}
  =\frac{n!}{T^{n}}.
\end{align}

From Fig.~\ref{sub_MD_0n}, it can be observed that the first PAoI of the monitor  is $A_1=T+X_1+X_2$, and $A_{i}=X_{i}+X_{i+1}(i\ge2)$.
Hence, the mean of the first PAoI $A_1$ is given as
\begin{align}\label{0n_A1}
  \mathbb{E}\left[A_1|(0,n)\right]&=\!\int_{0}^{T}\!\!\int_{0}^{T-x_{1}}\!\!\cdots\!\!\int_{0}^{T-\sum_{i=1}^{n-1}x_{i}}
  \!\!\!\!f_{X_{1},...,X_{n}}(x_{1},...,x_{n}|N(T)=n)(T\!+\!x_1\!+\!x_2){\rm d}x_{n}\cdots {\rm d}x_{2}{\rm d}x_{1}\nonumber\\
  &=\frac{n!}{T^{n}}\int_{0}^{T}\int_{0}^{T-x_{1}}\cdots\int_{0}^{T-\sum_{i=1}^{n-1}x_{i}}(T+x_{1}+x_{2}){\rm d}x_{n}\cdots {\rm d}x_{2}{\rm d}x_{1}.
\end{align}
To compute the above integration, we introduce the following lemma.
\begin{lem}\label{lem_X}
Let ${X_{1},X_{2},\cdots,X_{n}}$ be $n$ non-negative independent random variables satisfying $T-\sum_{i=1}^{n-1}X_i>0$. Then, the following equation holds
\begin{align}
  \int_{0}^{T-x_{2}-x_{1}}\cdots\int_{0}^{T-\sum_{i=1}^{n-1}x_{i}}1{\rm d}x_{n}\cdots {\rm d}x_{3}
  =\frac{(T-x_2-x_1)^{n-2}}{(n-2)!}.\nonumber
\end{align}
\begin{proof}
See Appendix \ref{prf:lem_x}.
\end{proof}
\end{lem}

Using Lemma \ref{lem_X}, we can rewrite (\ref{0n_A1}) as
\begin{align}
\mathbb{E}\left[A_1|(0,n)\right]=\frac{n!}{T^{n}}\int_{0}^{T}\int_{0}^{T-x_{1}}\frac{(T-x_2-x_1)^{n-2}}{(n-2)!}
 (T+x_{1}+x_{2}){\rm d}x_{2}{\rm d}x_{1}
 =\frac{(3+n)T}{1+n}.
\end{align}
The mean of the second PAoI $A_2$ can be expressed as
\begin{align}\label{0n_A}
  \mathbb{E}\left[A_2|(0,n)\right]&=\!\int_{0}^{T}\!\!\int_{0}^{T-x_{1}}\!\!\cdots\!\!\int_{0}^{T-\sum_{i=1}^{n-1}x_{i}}
  \!\!f_{X_{1},...,X_{n}}(x_{1},...,x_{n}|N(T)=n)(x_2+x_3){\rm d}x_{n}\cdots {\rm d}x_{2}{\rm d}x_{1}\nonumber\\
  &=\frac{n!}{T^{n}}\int_{0}^{T}\int_{0}^{T-x_{1}}\!\!\int_{0}^{T-x_{1}-x_{2}}\frac{(T-x_3-x_2-x_1)^{n-2}}{(n-2)!}
  (x_{2}+x_{3}){\rm d}x_3{\rm d}x_2{\rm d}x_1
  =\frac{2T}{1+n}.
\end{align}
Similarly, we have $\mathbb{E}\left[A_{n-1}|(0,n)\right]=\mathbb{E}\left[A_{n-2}|(0,n)\right]=\cdots=\mathbb{E}\left[A_2|(0,n)\right]$.
%\begin{align}
%  \mathbb{E}\left[A_2|(0,n)\right]=\mathbb{E}\left[A_3|(0,n)\right]=\cdots=\mathbb{E}\left[A_{n-1}|(0,n)\right].
%\end{align}
%\begin{equation}
%  \mathbb{E}\left[A_2\right]=\mathbb{E}\left[A_3\right]=\cdots=\mathbb{E}\left[A_{n-1}\right].
%\end{equation}

Therefore,  the sum average of PAoI for the M-D system in state $(0,n)$  is
\begin{equation}
  \mathbb{E}\left[A|(0,n)\right]=\sum_{i=1}^{n-1}\mathbb{E}\left[A_{i}|(0,n)\right]=\frac{(3n-1)T}{1+n}.
\end{equation}

The average of the shaded trapezoid areas $Q_1$, $Q_2$, and $Q_3$ in Fig.~\ref{sub_MD_0n} are given respectively by the following
\begin{align}
  \mathbb{E}\left[Q_1|(0,n)\right]&=\!\int_{0}^{T}\!\!\int_{0}^{T-x_{1}}\!\!\cdots\!\!\int_{0}^{T-\sum_{i=1}^{n-1}x_{i}}
   \!\!\!f_{X_{1},...,X_{n}}(x_{1},...,x_{n}|N(T)=n)\frac{(2T\!+\!x_1)x_1}{2}{\rm d}x_{n}\cdots {\rm d}x_{2}{\rm d}x_{1}\nonumber\\
  &=\frac{n!}{T^{n}}\int_{0}^{T}\frac{(T-x_1)^{n-1}}{(n-1)!}\cdot\frac{(2T+x_1)x_1}{2}{\rm d}x_1=\frac{(3+n)T^2}{(1+n)(2+n)},\\
  \mathbb{E}\left[Q_2|(0,n)\right]&=\int_{0}^{T}\int_{0}^{T-x_{1}}\cdots\int_{0}^{T-\sum_{i=1}^{n-1}x_{i}}f_{X_{1},...,X_{n}}(x_{1},...,x_{n}|N(T)=n)\nonumber\\
  &\qquad\times \frac{(2(T+x_1)+x_2)x_2}{2}{\rm d}x_{n}\cdots d{\rm d}x_{2}{\rm d}x_{1}
  =\frac{(4+n)T^2}{(1+n)(2+n)},\\
 \mathbb{E}\left[Q_3|(0,n)\right]&=\int_{0}^{T}\int_{0}^{T-x_{1}}\cdots\int_{0}^{T-\sum_{i=1}^{n-1}x_{i}}
f_{X_{1},...,X_{n}}(x_{1},...,x_{n}|N(T)=n)\nonumber\\
 \label{0n_Q} &\qquad\times \frac{(2x_2+x_3)x_3}{2}{\rm d}x_{n}\cdots {\rm d}x_{2}{\rm d}x_{1}=\frac{2T^2}{(1+n)(2+n)}.
\end{align}
Likewise, we obtain that $\mathbb{E}\left[Q_{n+1}|(0,n)\right]=\mathbb{E}\left[Q_n|(0,n)\right]=\cdots=\mathbb{E}\left[Q_3|(0,n)\right].$

Therefore, the mean area covered by the AoI waveform of the M-D system in state ${(0,n)}$ is
\begin{equation}
  \mathbb{E}\left[Q|(0,n)\right]=\sum_{i=1}^{n+1}\mathbb{E}\left[Q_i|(0,n)\right]=\frac{(4n+5)T^2}{(1+n)(2+n)}.
\end{equation}

\subsubsection{State $(k,0)$}
Sensor A sent $k(k\ge1)$ updates in the previous service period, yet it did not send new updates in the current period. In this case, the instantaneous AoI curve for the M-D system is shown in Fig.~\ref{sub_MD_k0}, where $S_{k}$  represents the time consumed by sensor A to send the $k\text{th}$ update in the previous period and $S_{k+1}$ represents the remaining time after sensor A successfully sent $k$ updates within a period $T$.
Based on the properties of Poisson process, the joint probability density function of $S_{k}$ and $S_{k+1}$ can be calculated as
\begin{align}\label{x_1x0}
  f_{S_{k},S_{k+1}}(s_{k},s_{k+1}|\widetilde{N}(T)=k)
  &=\frac{\mu e^{-\mu s_{k}}\mu e^{-\mu s_{k+1}}\frac{[\mu(T-s_{k}-s_{k+1})]^{k-2}}{(k-2)!}e^{-\mu(T-s_{k}-s_{k+1})}}{\frac{(\mu T)^{k}}{k!}e^{-\mu T}}\nonumber\\
  &=\frac{k(k-1)(T-s_{k}-s_{k+1})^{k-2}}{T^k}.
\end{align}
Then, the mean of the PAoI is
\begin{align}
 \!\mathbb{E}\left[A|(k,0)\right]\!=\!\!\int_{0}^{T}\!\!\int_{0}^{T-s_{k}}f_{S_{k},S_{k+1}}(s_{k},s_{k+1}|\widetilde{N}(T)\!=\!k)
 \left(s_{k}+s_{k+1}+T\right){\rm d}s_{k+1}{\rm d}s_{k}
 \!=\!\frac{(3\!+\!k)T}{1\!+\!k}.
\end{align}
The average of the area covered by the AoI waveform at the monitor is
\begin{align}
 \!\!\mathbb{E}\left[Q|(k,0)\right]\!=\!\!\int_{0}^{T}\!\!\int_{0}^{T\!-\!s_{k}}\!\!f_{S_{k},S_{k\!+\!1}}(s_{k},\!s_{k\!+\!1}|\widetilde{N}(T)\!=\!k)
 \frac{(2(s_{k}\!+\!s_{k+1})\!+\!T)T}{2}{\rm d}s_{k\!+\!1}{\rm d}s_{k}
 \!=\!\frac{(5\!+\!k)T^2}{2(1\!+\!k)}.
\end{align}

\subsubsection{State $(k,1)$}
Sensor A sent $k(k\ge1)$ updates in the previous service period, and transmitted one update in the current period, as shown in Fig.~\ref{sub_MD_k1}.
In the current service period, the monitor  received two valid updates, one from  sensor A and the other from sensor B.
We use the variable $X_{n}$ $(n={1,2})$ to represent the service time of the $n\text{th}$ update received by monitor in the current period.
Then, the conditional probability density function of $X_{1}$ is given by
\begin{align}\label{cpdf_x1}
  f_{X_{1}}(x_{1}|N(T)=1)=\frac{\mu e^{-\mu x_{1}}e^{-\mu(T-x_{1})}}{\mu T e^{-\mu T}}
  =\frac{1}{T}.
\end{align}

From Fig.~\ref{sub_MD_k1}, we observe that the first PAoI of the monitor before receiving an update from sensor A is  $A_1=S_k+S_{k+1}+X_{1}$, and the second PAoI  $A_2=S_{k+1}+T$.
Combining (\ref{x_1x0}) and (\ref{cpdf_x1}), the mean of the first PAoI $A_1$   is given by
\begin{align}\nonumber
  \mathbb{E}\left[A_{1}|(k,1)\right]&=\!\int_{0}^{T}\!\int_{0}^{T}\!\int_{0}^{T\!-\!s_{k}}\!f_{S_{k},S_{k+1}}\!(s_{k},s_{k+1}|\widetilde{N}(T)\!=\!k)\!
  f_{X_{1}}(x_{1}|N(T)\!=\!1)\\
  \label{k1_A1}&\qquad\times(s_{k}\!+\!s_{k+1}\!+\!x_{1}){\rm d}s_{k+1}{\rm d}s_{k}{\rm d}x_{1}=\frac{(5+k)T}{2(1+k)}.
\end{align}
The conditional probability density function of $S_{k+1}$ is given by
\begin{align}\label{cpdf_s_k+1}
  f_{S_{k+1}}(s_{k+1}|\widetilde{N}(T)=k)
  =\frac{\mu e^{-\mu s_{k+1}}\frac{[\mu(T-s_{k+1})]^{k-1}}{(k-1)!}e^{-\mu(T-s_{k+1})}}{\frac{(\mu T)^{k}}{k!}e^{-\mu T}}
  =\frac{k(T-s_{k+1})^{k-1}}{T^{k}}.
\end{align}
Therefore, the mean of the second PAoI $A_2$  can be expressed as
\begin{align}\label{k1_A2}
  \mathbb{E}\left[A_{2}|(k,1)\right]=\int_{0}^{T}f_{S_{k+1}}(s_{k+1}|\widetilde{N}(T)=k)(T+s_{k+1}){\rm d}s_{k+1}
  =\frac{(2+k)T}{(1+k)}.
\end{align}
Hence, we can obtain the sum of the average PAoI in this cases as
\begin{equation}\label{k1_A}
  \mathbb{E}\left[A|(k,1)\right]=\mathbb{E}\left[A_{1}|(k,1)\right]+\mathbb{E}\left[A_{2}|(k,1)\right]=\frac{(9+3k)T}{2(1+k)}.
\end{equation}

The average of the colored trapezoid areas $Q_1$ and $Q_2$ in Fig.~\ref{sub_MD_k1}  can be computed as
\begin{align}
  \mathbb{E}\left[Q_{1}|(k,1)\right]&=\int_{0}^{T}\int_{0}^{T}\int_{0}^{T-s_{k}}f_{S_{k},S_{k+1}}(s_{k},s_{k+1}|\widetilde{N}(T)=k)\nonumber\\
  &\quad\quad \times f_{X_{1}}(x_{1}|N(T)=1)\frac{(2(s_{k}+s_{k+1})+x_{1})x_{1}}{2}{\rm d}s_{k+1}{\rm d}s_{k}{\rm d}x_{1}
  \label{k1_Q1}=\frac{(7+k)T^2}{6(1+k)},\\
  \mathbb{E}\left[Q_{2}|(k,1)\right]&=\int_{0}^{T}\int_{0}^{T}f_{S_{k+1}}(s_{k+1}|\widetilde{N}(T)=k)f_{X_{1}}(x_{1}|N(T)=1)\nonumber\\
  &\quad\quad\times\frac{(s_{k+1}+x_{1}+s_{k+1}+T)(T-x_{1})}{2}{\rm d}s_{k+1}{\rm d}x_{1}
  \label{k1_Q2}=\frac{(5+2k)T^2}{6(1+k)}.
\end{align}
Therefore, the average area under the instantaneous AoI curve  of the M-D system in state $(k,1)$ is given by
\begin{equation}\label{k1_Q}
  \mathbb{E}\left[Q|(k,1)\right]=\mathbb{E}\left[Q_{1}|(k,1)\right]+\mathbb{E}\left[Q_{2}|(k,1)\right]=\frac{(4+k)T^2}{2(1+k)}.
\end{equation}

\subsubsection{State $(k,n)$ }
Sensor A transmitted $k$ ($k\ge1$) updates in the previous service period, and transmits $n$ $(n\ge2)$ updates in the current period, as shown in Fig.~\ref{sub_MD_kn}.
From this figure, we can conclude that the first PAoI is $A_1=S_k+S_{k+1}+X_{1}$, the second PAoI is $A_2=S_{k+1}+X_{1}+X_{2}$, and the $i\text{th}$ PAoI is $A_{i}=X_{i-1}+X_{i}(i=3,\cdots,n)$.
Combining (\ref{0n_jpdf}), (\ref{x_1x0}), and Lemma \ref{lem_X}, the mean of the first PAoI $A_1$ is given by
\begin{align}
  \mathbb{E}\left[A_1|(k,n)\right]&=\int_{0}^{T}\int_{0}^{T-x_{1}}\!\cdots\!\int_{0}^{T-\sum_{i=1}^{n-1}x_{i}}\int_{0}^{T}\int_{0}^{T-s_{k}}
  f_{X_{1},...,X_{n}}(x_{1},...,x_{n}|N(T)=n)\nonumber\\
  &\quad \times f_{S_{k},S_{k+1}}(s_{k},s_{k+1}|\widetilde{N}(T)=k)
  (s_{k+1}+s_{k}+x_{1}){\rm d}s_{k+1}{\rm d}s_{k}{\rm d}x_{n}\cdots {\rm d}x_{2}{\rm d}x_{1}\nonumber\\
  &=\frac{n!}{T^{n}}\int_{0}^{T}\int_{0}^{T-x_{1}}\!\cdots\!\int_{0}^{T-\sum_{i=1}^{n-1}x_{i}}
  \left(\frac{2T}{1+k}+x_1\right){\rm d}x_{n}\cdots {\rm d}x_{2}{\rm d}x_{1}
  =\frac{(3+k+2n)T}{(1+k)(1+n)}.
\end{align}
Combining (\ref{0n_jpdf}), (\ref{cpdf_s_k+1}), and Lemma \ref{lem_X}, the mean of the second PAoI $A_2$ is given by
\begin{align}
  \mathbb{E}\left[A_2|(k,n)\right]&=\int_{0}^{T}\int_{0}^{T-x_{1}}\!\cdots\!\int_{0}^{T-\sum_{i=1}^{n-1}x_{i}}\int_{0}^{T}
  f_{X_{1},...,X_{n}}(x_{1},...,x_{n}|N(T)=n)\nonumber\\
  &\quad\times f_{S_{k+1}}(s_{k+1}|\widetilde{N}(T)=k)
  (s_{k+1}+x_{1}+x_{2}){\rm d}s_{k+1}{\rm d}x_{n}\!\cdots\! {\rm d}x_{2}{\rm d}x_{1}
  =\frac{(3+2k+n)T}{(1+k)(1+n)}.
\end{align}
The derivation of the remaining $(n-2)$ PAoI is the same as that of (\ref{0n_A}).
Therefore,  the sum average of PAoI for the M-D system in state $(k,n)$ is
\begin{equation}
  \mathbb{E}\left[A|(k,n)\right]=\sum_{i=1}^{n}\mathbb{E}\left[A_i|(k,n)\right]=\frac{(2nk+5n-k+2)T}{(1+k)(1+n)}.
\end{equation}

Combining (\ref{0n_jpdf}), (\ref{x_1x0}), and Lemma \ref{lem_X}, the average of the shaded trapezoid area $Q_1$  is
\begin{align}
  \mathbb{E}\left[Q_1|(k,n)\right]&=\int_{0}^{T}\cdots\int_{0}^{T-\sum_{i=1}^{n-1}x_{i}}\int_{0}^{T}\int_{0}^{T-s_{k}}
  f_{X_{1},...,X_{n}}(x_{1},...,x_{n}|N(T)=n)\nonumber\\
  &\quad\times f_{S_{k},S_{k+1}}(s_{k},s_{k+1}|\widetilde{N}(T)=k)
  \frac{(2(s_{k+1}+s_{k})+x_1)x_1}{2}{\rm d}s_{k+1}{\rm d}s_{k}{\rm d}x_{n}\cdots {\rm d}x_{1}\nonumber\\
  &=\frac{(2n+k+5)T^2}{(1+k)(1+n)(2+n)}.
\end{align}
Combining (\ref{0n_jpdf}), (\ref{cpdf_s_k+1}), and Lemma \ref{lem_X}, the average of the shaded trapezoid area $Q_2$ is
\begin{align}
  \mathbb{E}\left[Q_2|(k,n)\right]&=\int_{0}^{T}\int_{0}^{T-x_{1}}\cdots\int_{0}^{T-\sum_{i=1}^{n-1}x_{i}}\int_{0}^{T}
  f_{X_{1},...,X_{n}}(x_{1},...,x_{n}|N(T)=n)\nonumber\\
  &\quad\times f_{S_{k+1}}(s_{k+1}|\widetilde{N}(T)=k)
  \frac{(2(s_{k+1}+x_1)+x_2)}{2}x_2{\rm d}s_{k+1}{\rm d}x_{n}\cdots {\rm d}x_{2}{\rm d}x_{1}\nonumber\\
  &=\frac{(n+2k+4)T^2}{(1+k)(1+n)(2+n)}.
\end{align}
The derivation of the remaining $(n-1)$ trapezoids is the same as that of (\ref{0n_Q}).
Hence, we have the average area covered by AoI waveform of the M-D system in state ${(k,n)}$ given as
\begin{equation}
  \mathbb{E}\left[Q|(k,n)\right]=\sum_{i=1}^{n+1}\mathbb{E}\left[Q_i|(k,n)\right]=\frac{(2nk+5n+k+7)T^2}{(1+k)(1+n)(2+n)}.
\end{equation}

\subsection{Average PAoI of the M-D System}\label{proof_md_paoi}
\allowdisplaybreaks
In a service period, the average PAoI of the M-D system can be obtained by dividing the sum of PAoI by the number of PAoI.
Table \ref{table_A} summarizes the number of PAoI and the sum of PAoI in the current service period for the M-D system in different states.
\begin{table}
  \centering
  \caption{The number of PAoI $\mathbb{E}\left[N|(k,n)\right]$ and the sum average of PAoI $\mathbb{E}\left[A|(k,n)\right]$ in the current period.}\label{table_A}
 \renewcommand\arraystretch{1.5}
  \begin{tabular}{!{\vrule width1.2pt}c|p{1.5cm}<{\centering}|p{1.5cm}<{\centering}|p{1.5cm}<{\centering}||c|p{1.5cm}<{\centering}|p{1.5cm}<{\centering}|p{2.1cm}<{\centering}!{\vrule width1.2pt}}
 \Xhline{1.2pt}
  \multicolumn{4}{!{\vrule width1.2pt}c||}{$\bm{k=0}$} & \multicolumn{4}{c!{\vrule width1.2pt}}{$\bm{k\ge1}$} \\
 \hline
 & $\bm{n=0}$ & $\bm{n=1}$ & $\bm{n\ge2}$ & & $\bm{n=0}$ & $\bm{n=1}$ & $\bm{n\ge2}$ \\
 \hline
 \cellcolor{blue!10}$\mathbb{E}\left[A|(k,n)\right]$ & \cellcolor{blue!10}$2T$ & \cellcolor{blue!10}$2T$   &  \cellcolor{blue!10} $\frac{(3n-1)T}{1+n}$
 & \cellcolor{blue!10}$\mathbb{E}\left[A|(k,n)\right]$ & \cellcolor{blue!10} $\frac{(3+k)T}{1+k}$ & \cellcolor{blue!10} $\frac{(9+3k)T}{2(1+k)}$   &\cellcolor{blue!10}$\frac{(2nk+5n-k+2)T}{(1+k)(1+n)}$\\
 \hline
 $\mathbb{E}\left[N|(k,n)\right]$ & 1 & 1 & $n-1$ &$\mathbb{E}\left[N|(k,n)\right]$ & 1 & 2 & $n$\\
\Xhline{1.2pt}
  \end{tabular}
  \vspace{-1.5em}
\end{table}

%\begin{table}
%  \centering
%  \caption{The number of peak ages and the sum of peak ages in the current period}\label{table_A}
% \renewcommand\arraystretch{1.5}
%  \begin{tabular}{|l|c|c|c|c|}
%  \hline
%  \multicolumn{2}{|c|}{} & $\bm{n=0}$ & $\bm{n=1}$ & $\bm{n\ge2}$ \\
% \hline
%  \multirow{2}{*}{$\bm{k=0}$} & \cellcolor{blue!20}$A$ & \cellcolor{blue!20}$2T$ & \cellcolor{blue!20}$2T$   &  \cellcolor{blue!20} $\frac{(3n-1)T}{1+n}$ \\
%  \cline{2-5}
%  &$N(\tau)$ & 1 & 1 & $n-1$\\
%  \hline
%  \multirow{2}{*}{$\bm{k\ge1}$} & \cellcolor{blue!20}$A$ & \cellcolor{blue!20} $\frac{(3+k)T}{1+k}$ & \cellcolor{blue!20} $\frac{(9+3k)T}{2(1+k)}$   &  \cellcolor{blue!20} $\frac{(2nk+5n-k+2)T}{(1+k)(1+n)}$ \\
%  \cline{2-5}
%  &$N(\tau)$ & 1 & 2 & $n$\\
%  \hline
%  \end{tabular}
%\end{table}

According to Table \ref{table_A} and  formula  (\ref{possion}), the average number of PAoI of the M-D system in a period can be calculated by
\begin{align}\label{total_num}
\mathbb{E}\left[N\right]&=\mathrm{Pr}\{\widetilde{N}(T)=0\}\Big[\mathrm{Pr}\{N(T)=0\}+\mathrm{Pr}\{N(T)=1\}\Big]\nonumber\\
&\quad+\mathrm{Pr}\{\widetilde{N}(T)\!=\!0\}\left[\sum_{n=2}^{\infty}\mathrm{Pr}\{N(T)\!=\!n\}(n\!-\!1)\right]
+\mathrm{Pr}\{N(T)\!=\!0\}\left[\sum_{k=1}^{\infty}\mathrm{Pr}\{\widetilde{N}(T)\!=\!k\}\right]\nonumber\\
&\quad+\mathrm{Pr}\{N(T)\!=\!1\}\left[\sum_{k=1}^{\infty}\mathrm{Pr}\{\widetilde{N}(T)\!=\!k\}\times \!2\right]
+\sum_{k=1}^{\infty}\sum_{n=2}^{\infty}\Big[\mathrm{Pr}\{\widetilde{N}(T)\!=\!k\}\mathrm{Pr}\{N(T)\!=\!n\}n\Big]\nonumber\\
&=e^{-2\mu T}+\mu Te^{\mu T}+\mu T.
\end{align}
The sum of the average PAoI for the M-D system is calculated as
\begin{align}\label{total_A}
\mathbb{E}\left[A\right]&=\mathrm{Pr}\{\widetilde{N}(T)=0\}\mathrm{Pr}\{N(T)=0\}\times 2T
+\mathrm{Pr}\{\widetilde{N}(T)=0\}\mathrm{Pr}\{N(T)=1\}\times 2T\nonumber\\
&\quad+\mathrm{Pr}\{\widetilde{N}(T)=0\}\left[\sum_{n=2}^{\infty}\mathrm{Pr}\{N(T)=n\}\frac{(3n-1)T}{1+n}\right]\nonumber\\
&\quad+\mathrm{Pr}\{N(T)=0\}\left[\sum_{k=1}^{\infty}\mathrm{Pr}\{\widetilde{N}(T)=k\}\frac{(3+k)T}{1+k}\right]\nonumber\\
&\quad+\mathrm{Pr}\{N(T)=1\}\left[\sum_{k=1}^{\infty}\mathrm{Pr}\{\widetilde{N}(T)=k\}\frac{(9+3k)T}{2(1+k)}\right]\nonumber\\
&\quad+\sum_{k=1}^{\infty}\sum_{n=2}^{\infty}\left[\mathrm{Pr}\{\widetilde{N}(T)=k\}\mathrm{Pr}\{N(T)=n\} \times \frac{(2nk+5n-k+2)T}{(1+k)(1+n)}\right]\nonumber\\
&=\frac{e^{-2\mu T}\left[2+2\mu T+e^{\mu T}(-2+\mu T(2e^{\mu T}+\mu T))\right]}{\mu}.
\end{align}
Therefore, the average PAoI of the M-D system can be calculated by
\begin{equation}\label{MD_PAoI_yuanshi}
\Delta_\text{M-D}^\text{(peak)}=\frac{\mathbb{E}\left[A\right]}{\mathbb{E}\left[N\right]}.
\end{equation}
Substituting (\ref{total_num}) and (\ref{total_A}) into (\ref{MD_PAoI_yuanshi}), it yields (\ref{MD_PAoI}), which completes the proof of Proposition \ref{prop_md_paoi}.

\subsection{Average AoI of the M-D System}\label{proof_md_aoi}
\allowdisplaybreaks
Based on the probability of the M-D system in each state, the average AoI of the M-D system can be obtained by weighting and summing the average AoI of each state.
In a service period, the average AoI of the M-D system in each state can be obtained by dividing the area covered by AoI waveform $Q$ by time interval $T$.
Table \ref{table_Q} summarizes the average of the area covered by AoI waveform of the M-D system in different states.
\begin{table}
  \centering
  \caption{The average calculated area $\mathbb{E}\left[Q|(k,n)\right]$ covered by AoI waveform of the M-D system in different states.}\label{table_Q}
 \renewcommand\arraystretch{1.5}
 \normalsize
  \begin{tabular}{!{\vrule width1.2pt}l|p{2.5cm}<{\centering}|p{2.5cm}<{\centering}||l|p{2.5cm}<{\centering}|p{2.5cm}<{\centering}!{\vrule width1.2pt}}
  \Xhline{1.2pt}
  \multirow{2}{*}{$\bm{k=0}$} & $\bm{n=0}$ & $\bm{n\ge1}$ & \multirow{2}{*}{$\bm{k\geq1}$} &  $\bm{n=0}$ & $\bm{n\ge1}$\\
  \cline{2-3}
  \cline{5-6}
  & $\frac{3T^2}{2}$ & $\frac{(4n+5)T^2}{(1+n)(2+n)}$ & & $\frac{(5+k)T^2}{2(1+k)}$ & $\frac{(2nk+5n+k+7)T^2}{(1+k)(1+n)(2+n)}$\\
  \Xhline{1.2pt}
  \end{tabular}
  \vspace{-1.5em}
\end{table}

%\begin{table}
%  \centering
%  \caption{The average calculated area $Q$ covered by AoI waveform of the M-D system in different states}\label{table_Q}
% \renewcommand\arraystretch{2}
% \normalsize
%  \begin{tabular}{|c|c|c|c|}
%  \hline
%  \multicolumn{2}{|c||}{$\bm{k=0}$} & \multicolumn{2}{c|}{$\bm{k\ge1}$} \\
% \hline
%  $\bm{n=0}$ & $\bm{n\ge1}$ &   $\bm{n=0}$ & $\bm{n\ge1}$\\
% \hline
%   $\frac{3T^2}{2}$ & $\frac{(4n+5)T^2}{(1+n)(2+n)}$  & $\frac{(5+k)T^2}{2(1+k)}$ & $\frac{(2nk+5n+k+7)T^2}{(1+k)(1+n)(2+n)}$\\
%  \hline
%  \end{tabular}
%\end{table}

Combining Table \ref{table_Q} and (\ref{possion}), we can get the  average AoI of the M-D system
\begin{align}\label{eq_MD_AoI}
  \Delta_\text{M-D}&=\mathrm{Pr}\{\widetilde{N}(T)\!=\!0\}\mathrm{Pr}\{N(T)\!=\!0\}\!\times\! \frac{3T}{2}
  +\mathrm{Pr}\{N(T)\!=\!0\}\sum_{k=1}^{\infty}\left[\mathrm{Pr}\{\widetilde{N}(T)\!=\!k\}\frac{(5+k)T}{2(1+k)}\right]\nonumber\\
  &\quad+\mathrm{Pr}\{\widetilde{N}(T)=0\}\sum_{n=1}^{\infty}\left[\mathrm{Pr}\{N(T)=n\}\frac{(4n+5)T}{(1+n)(2+n)}\right]\nonumber\\
  &\quad+\sum_{k=1}^{\infty}\sum_{n=2}^{\infty}\left[\mathrm{Pr}\{\widetilde{N}(T)=k\}\mathrm{Pr}\{N(T)=n\} \frac{(2nk+5n+k+7)T}{(1+k)(1+n)(2+n)}\right]\nonumber\\
  &=\frac{3+2T\mu+e^{T\mu}(-3+(-1+2e^{T\mu})T\mu)}{T\mu^{2}e^{2T\mu}}.
\end{align}
This completes the proof of Proposition \ref{prop_md_aoi}.
%\begin{rem}
%When $T=1/\mu$, the average AoI and pAoI of the M-D system are $(5e^{-2}-4e^{-1}+2)/\mu$ and $(5e^{2}-e+4)/[(e^2+e+1)\mu]$. Compared with the average AoI and pAoI of the single-source update system with service rate $\mu$ (see \cite[Sec. \uppercase\expandafter{\romannumeral7}]{ut_1}), both of which are $2/\mu$, the average AoI and pAoI of the M-D system are reduced by $\big(1-\frac{(5e^{-2}-4e^{-1}+2)/\mu}{2/\mu}\big)\times100\%\approx39.7\%$ and $\big(1-\frac{(5e^{2}-e+4)/[(e^2+e+1)\mu]}{2/\mu}\big)\times100\%\approx27.7\%$.
%\end{rem}
%\begin{rem}
%When $\mu\to0$, the average AoI $\Delta_\text{M-D}$ of the M-D system is $\lim_{\mu\to0}\Delta_\text{M-D}=3T/2$, which is equal to the average AoI of the M/D/1 non-preemptive queue with ZW  policy \cite[Sec. \uppercase\expandafter{\romannumeral7}]{ut_1}.
%\end{rem}

\section{Numerical Results}\label{section_results}
\allowdisplaybreaks
In this section, we provide numerical results to evaluate the performance of dual-queue updating systems.
%First, we carry out simulations to validate the theoretical analysis on the AoI of M-M and M-D systems.
%Then, we demonstrate the effectiveness of dual updating by comparing the AoI performance of the dual-queue system with the single-queue system.
% Finally, we compare the AoI performance of the M-M and M-D systems to gain better insights on the characteristics of deterministic service systems and the random service systems.
\begin{figure}
  \centering
\subfigure[Average PAoI]{\label{PAoI}\includegraphics[scale=0.54]{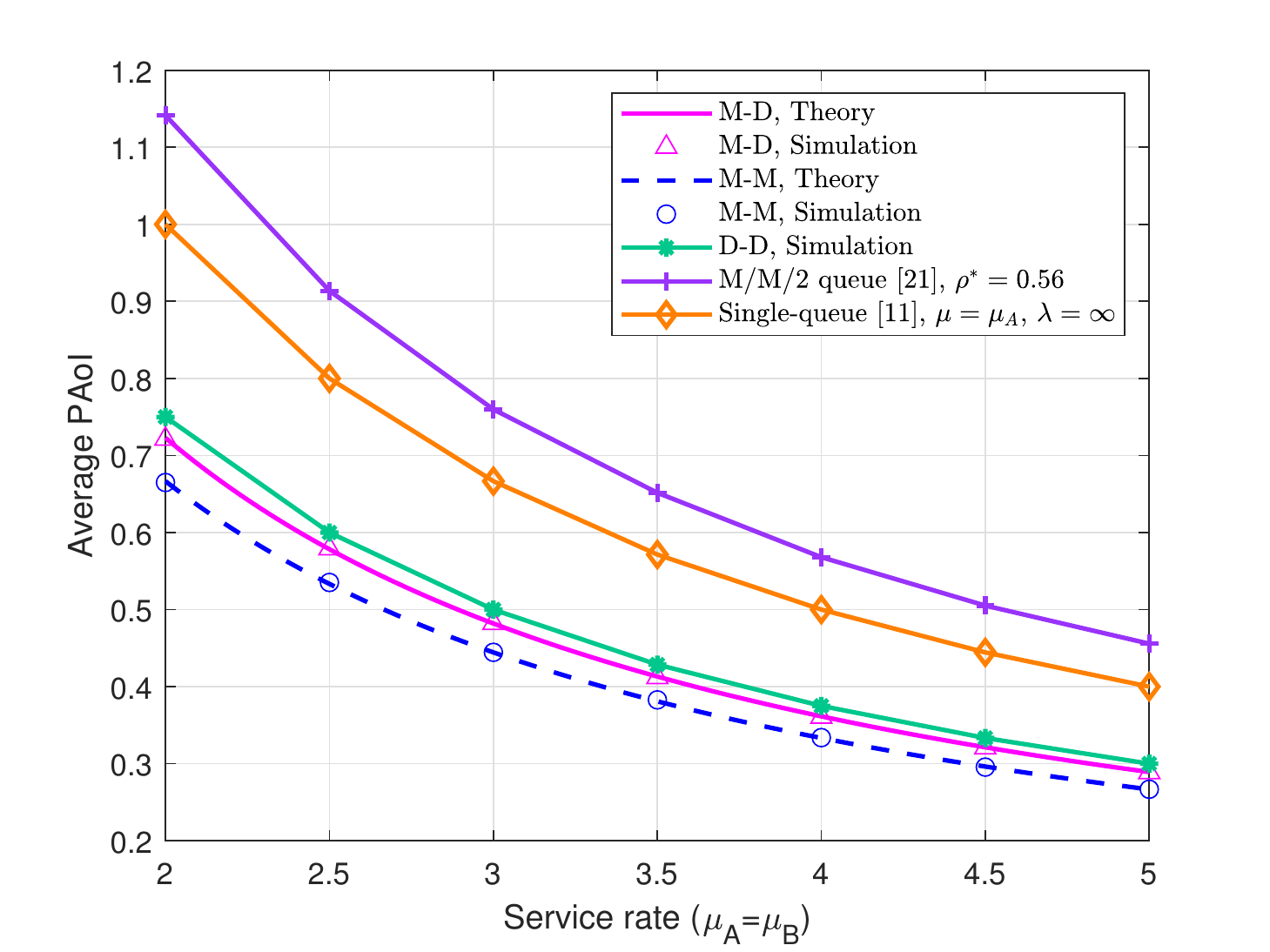}}\hspace{0.05in}
\subfigure[Average AoI]{\label{AoI}\includegraphics[scale=0.54]{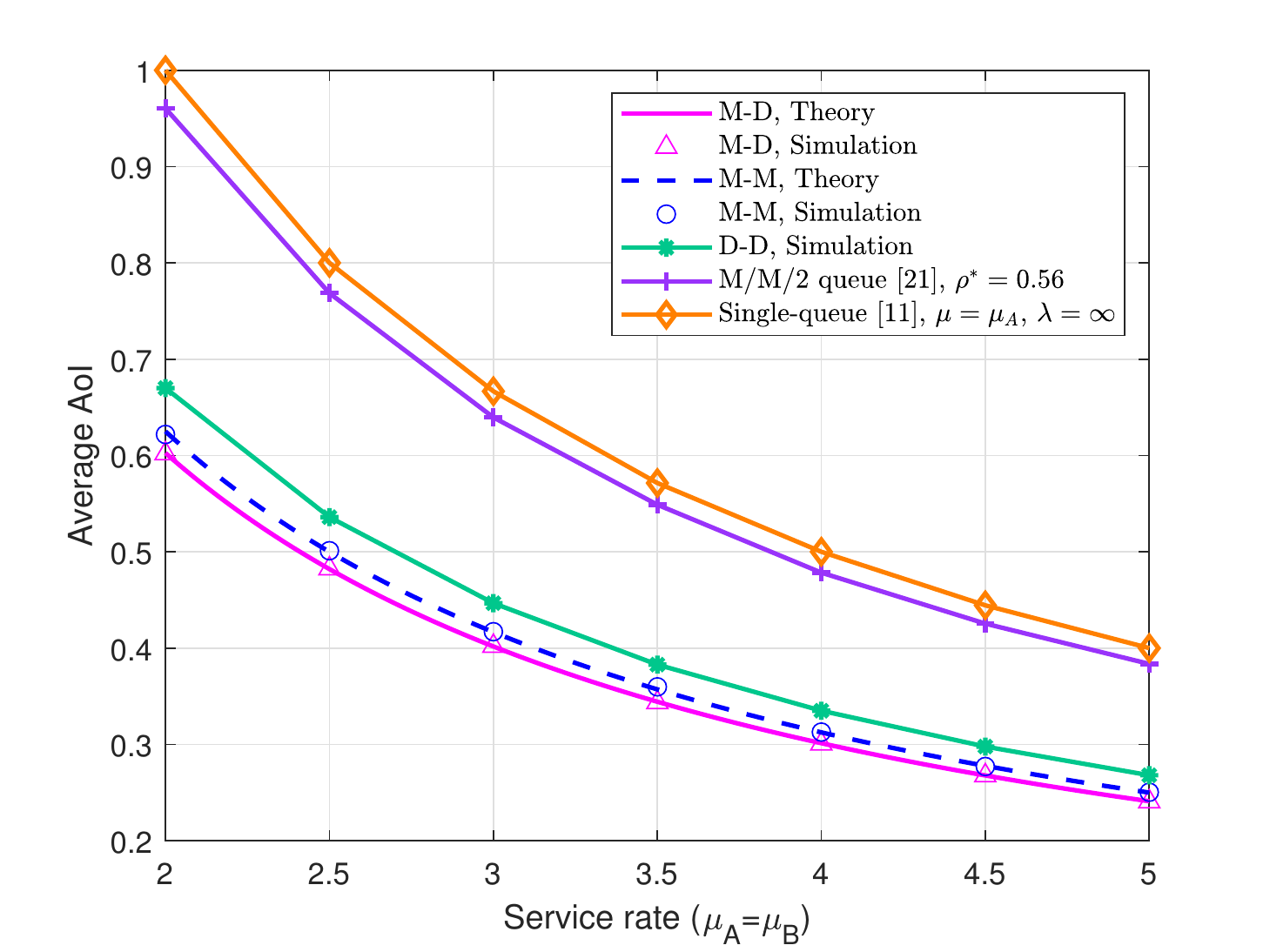}}
  \caption{The average PAoI and average AoI for the M-M system and the M-D system, where $\mu_A=\mu_B$.}
  \label{Fig_simu}
  \vspace{-1.5em}
\end{figure}
Based on the expressions derived in (\ref{MM_PAoI}) and (\ref{MD_PAoI}), Fig.~\ref{PAoI} depicts the average PAoI of M-M system and M-D system as the service rate increases from 2 to 5, where we set $\mu_A=\mu_B$. Using the same parameters, we simulate the M-M system and M-D system, respectively, and plot the corresponding results as markers in Fig.~\ref{PAoI}, where each point is the average PAoI of the considered systems serving $10^4$
updates. From this figure, we can see that the simulation results match well with the theoretical results, which validates the theoretical findings of this paper. Additionally, we take the single-queue  system with exponentially distributed service time, the D-D system where the service time of two sensors is fixed, and the M/M/2 queue for comparison. The load of the M/M/2 queue is set as the optimal load $\rho^{*}\!=\!\frac{\lambda}{\mu_{A}+\mu_{B}}\!=\!0.56$ (\cite{MM2_conference}). We can see that the M-M system which has the highest level of randomness in the service time performs the best in terms of the average PAoI, followed by the M-D system and the D-D system which possesses no randomness in the service time. As such, compared with the single-queue update system, the dual-queue update system can effectively reduce the average PAoI.

Fig.~\ref{AoI} plots the average AoI of M-M system and M-D system under different service rates, based on the the expressions given by (\ref{MM_AoI}) and (\ref{MD_AoI}). We set the service rate of sensor A and sensor B to the same value.  We observe that the simulations  coincide with the theoretical results, verifying the correctness of the derived average AoI expression. It is worth noting that the average AoI performance of the M/M/2 queue is worse than that of all dual-queue systems. This is because packets need to be queued for service  in the M/M/2 queue, resulting in a larger AoI. In contrast, in the dual-queue system considered in this work, updates will be transmitted directly once generated, thus maintaining freshness.

%Additionally, for single-queue systems, the level of randomness in the service time reciprocally affects the average AoI performance, i.e., the system with higher randomness of service time distribution has worse average AoI.
%However, this phenomenon does not occur in dual-queue update systems. As this experiment shows, the M-D system attains the best AoI performance.
%\begin{figure}[t]
%  \centering
%  \includegraphics[scale=0.65]{AoI}
%  \caption{The average AoI for M-M system and M-D system, where $\mu_A=\mu_B$. }
%  \label{AoI}
%\end{figure}

\begin{figure}[t]
  \centering
\subfigure[Average PAoI]{\label{MM_MD_PAoI_compare}\includegraphics[scale=0.54]{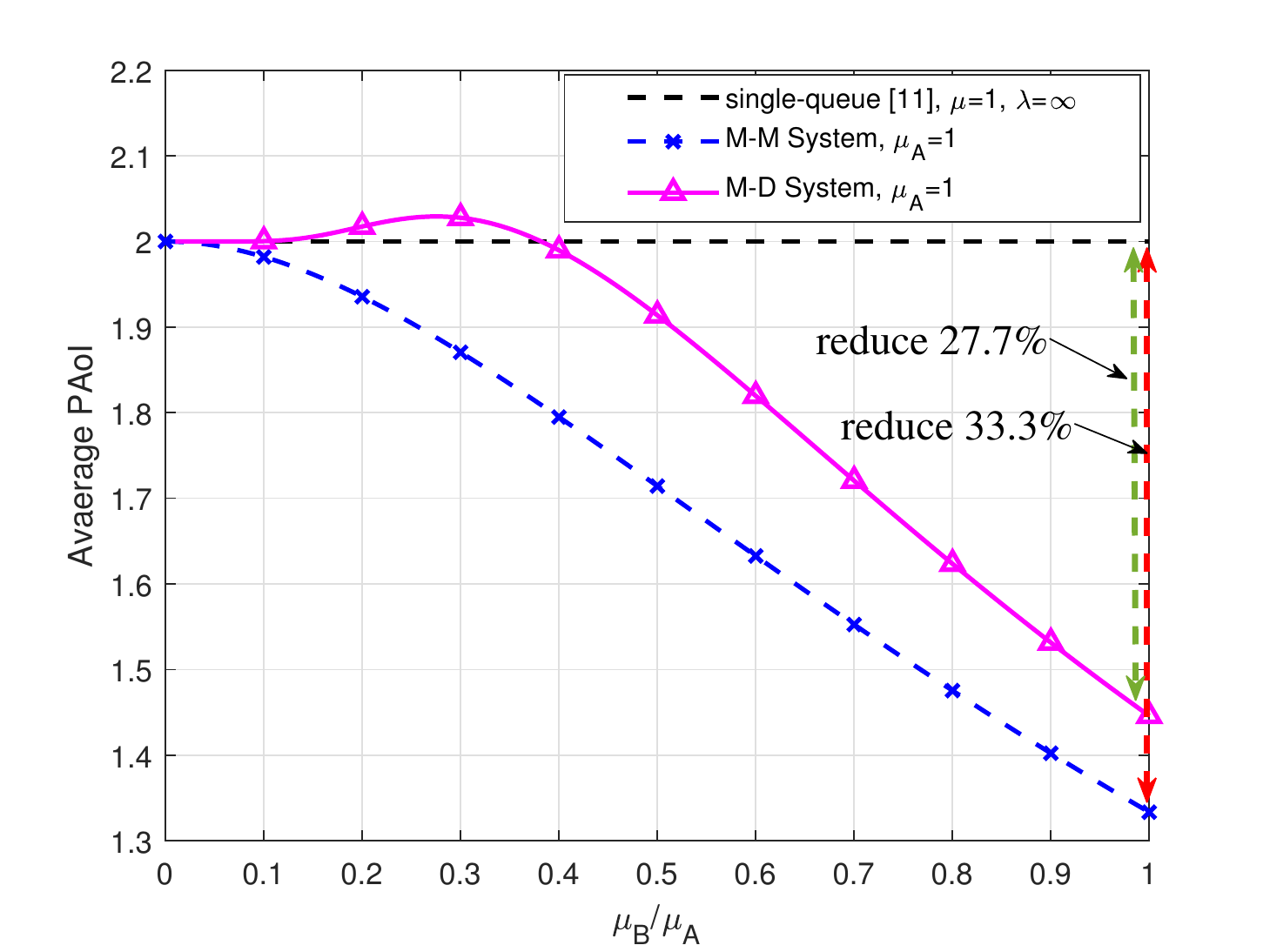}}\hspace{0.05in}
\subfigure[Average AoI]{\label{MM_MD_AoI_compare}\includegraphics[scale=0.54]{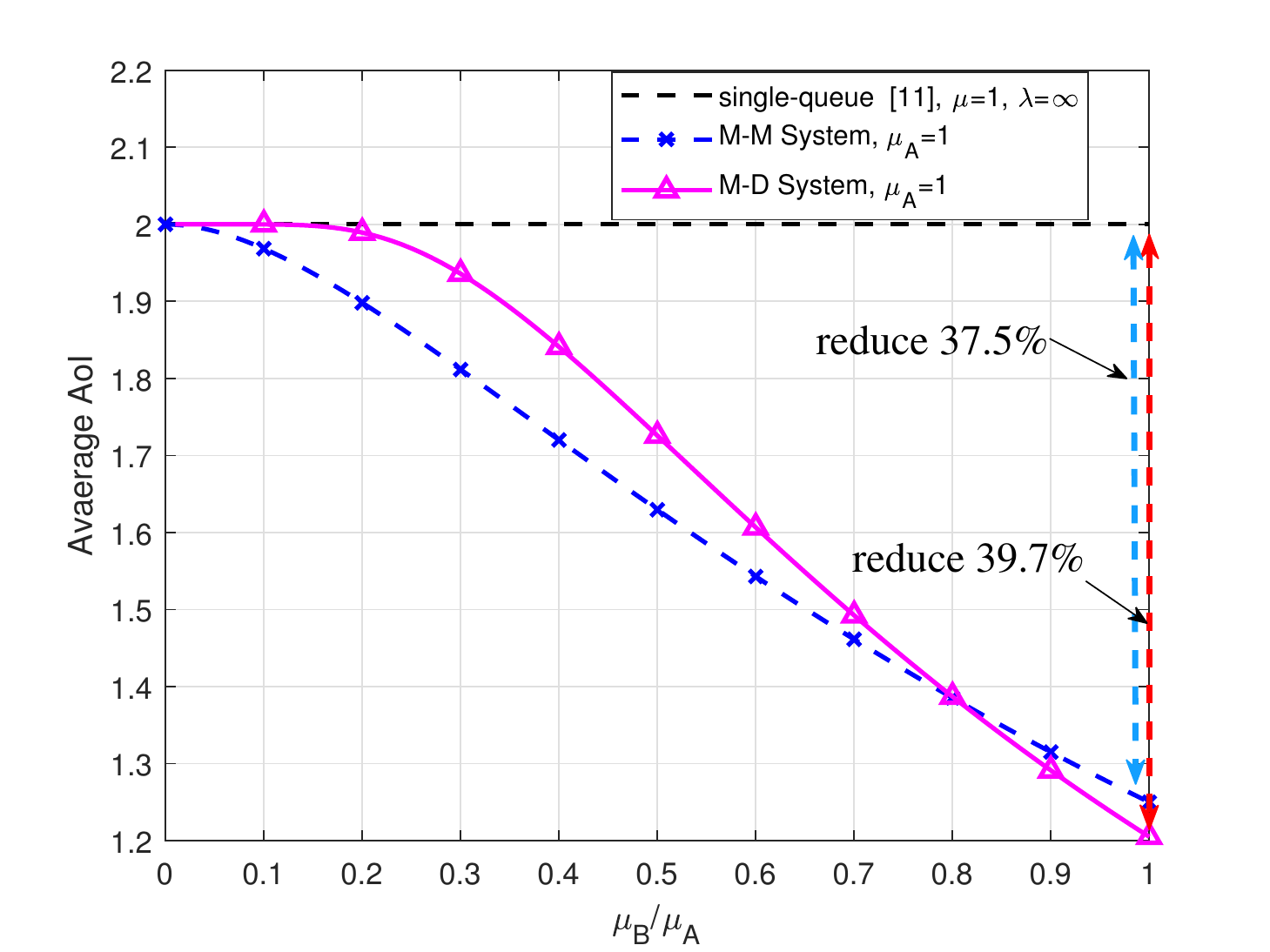}}
  \caption{Average PAoI and average AoI of the M-M system and the M-D system as the ratio of service rate  $\mu_B/\mu_A$ increasing from $0$ to $1$. The service rate of sensor A is set as $\mu_A=1$.}
  \label{Fig_compare}
  \vspace{-1.5em}
\end{figure}
Fig.~\ref{MM_MD_PAoI_compare} depicts the average PAoI of M-M system and M-D system as a function of the ratio of service rate $\mu_B/\mu_A$,  where the service rate of sensor A is fixed as $\mu_A=1$. This figure shows that the M-M system always outperforms M-D system in terms of average PAoI. Moreover, the average PAoI of M-M system decreases monotonically with the growth of service rate ratio $\mu_B/\mu_A$. In contrast, the average PAoI of M-D system  first increases and then decreases along with the rise of $\mu_B/\mu_A$.
This is because, when $\mu_B$ is small, with the increase of $\mu_B$, the proportion of updates received by the monitor from sensor B with fixed lager service time also increases, leading a slight increase in the average PAoI of the M-D system. When $\mu_B$ is large, the average AoI of the M-D system is mainly affected by the service rate, i.e., it decreases with the increase of the service rate.
We also take the single-queue update system in which the service time follows the exponential distribution with parameter $\mu=1$ and the generation rate is infinite (see \cite{ref_jit}) for comparison, and we draw it as the black dotted curve.
Note that when the service rates of sensor A and sensor B are the same,  the average PAoI of the M-M system and M-D system are reduced by $(2-1.333)/2\times 100\%=33.3\%$ and $(2-1.446)/2\times 100\%=27.7\%$ over the single-queue update system.
Therefore, in order to optimize the average PAoI performance of the considered system,  a sensor with memoryless service time needs to be added to the system, i.e., one shall employ the M-M system.

%
%\begin{figure}[t]
%  \centering
%  \includegraphics[scale=0.65]{MM_MD_PAoI_compare}
%  \caption{Average PAoI of M-M system and M-D system as the ratio of service rate  $\mu_B/\mu_A$ increasing from $0$ to $1$. The service rate of sensor A is set as $\mu_A=1$. }
%  \label{MM_MD_PAoI_compare}
%\end{figure}

Fig.~\ref{MM_MD_AoI_compare} illustrates the  average AoI of M-M system and M-D system as a function of the ratio of service rate $\mu_B/\mu_A$.
The black dotted line in Fig.~\ref{MM_MD_AoI_compare} represents the average AoI of a single-queue update system in which the service time follows an exponential distribution with parameter $\mu=1$ (see \cite{ref_jit}).
The service rate of sensor A is fixed as $\mu_A=1$.  As shown in Fig.~\ref{MM_MD_AoI_compare},  the average AoI of the M-M system and M-D system decays monotonically with an increase in the service rate ratio $\mu_B/\mu_A$.
Notably, there exists a critical point in the service rate ratio, i.e., $\mu_B/\mu_A=0.8$, below which the average AoI performance of the M-M system is better than that of the M-D system.
However, the M-D system outperforms M-M system when the service rate exceeds this critical point.
Because when $\mu_{B}$ is larger, the proportion of updates received by the monitor from sensor B is higher, and the improvement of AoI performance caused by the determined service time is more obvious.
Moreover,  the M-M system and M-D system can respectively reduce the average AoI by $(2-1.25)/2\times 100\%=37.5\%$ and $(2-1.205)/2\times 100\%=39.7\%$ over the single-queue update system.
%\begin{figure}[t]
%  \centering
%  \includegraphics[scale=0.65]{MM_MD_AoI_compare}
%  \caption{Average AoI of M-M system and M-D system as the ratio of service rate  $\mu_B/\mu_A$ increases from $0$ to $1$. The service rate of sensor A is set as $\mu_A=1$. }
%  \label{MM_MD_AoI_compare}
%\end{figure}

\begin{figure}[t]
  \centering
\subfigure[Effective arrive rate]{\label{fig_effective_rate}\includegraphics[scale=0.54]{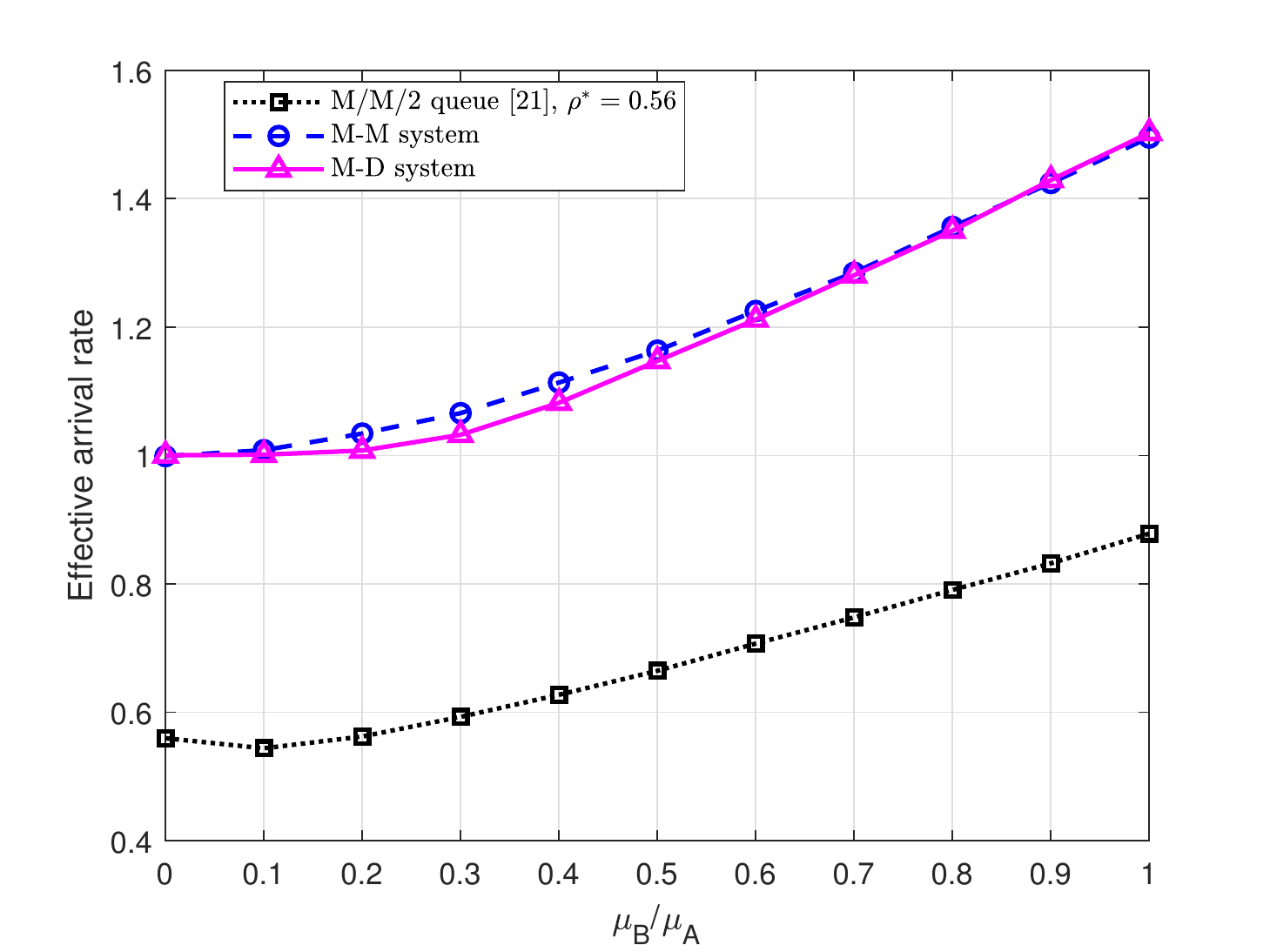}}
\subfigure[Obsolete packet ratio]{\label{fig_expired_rate}\includegraphics[scale=0.54]{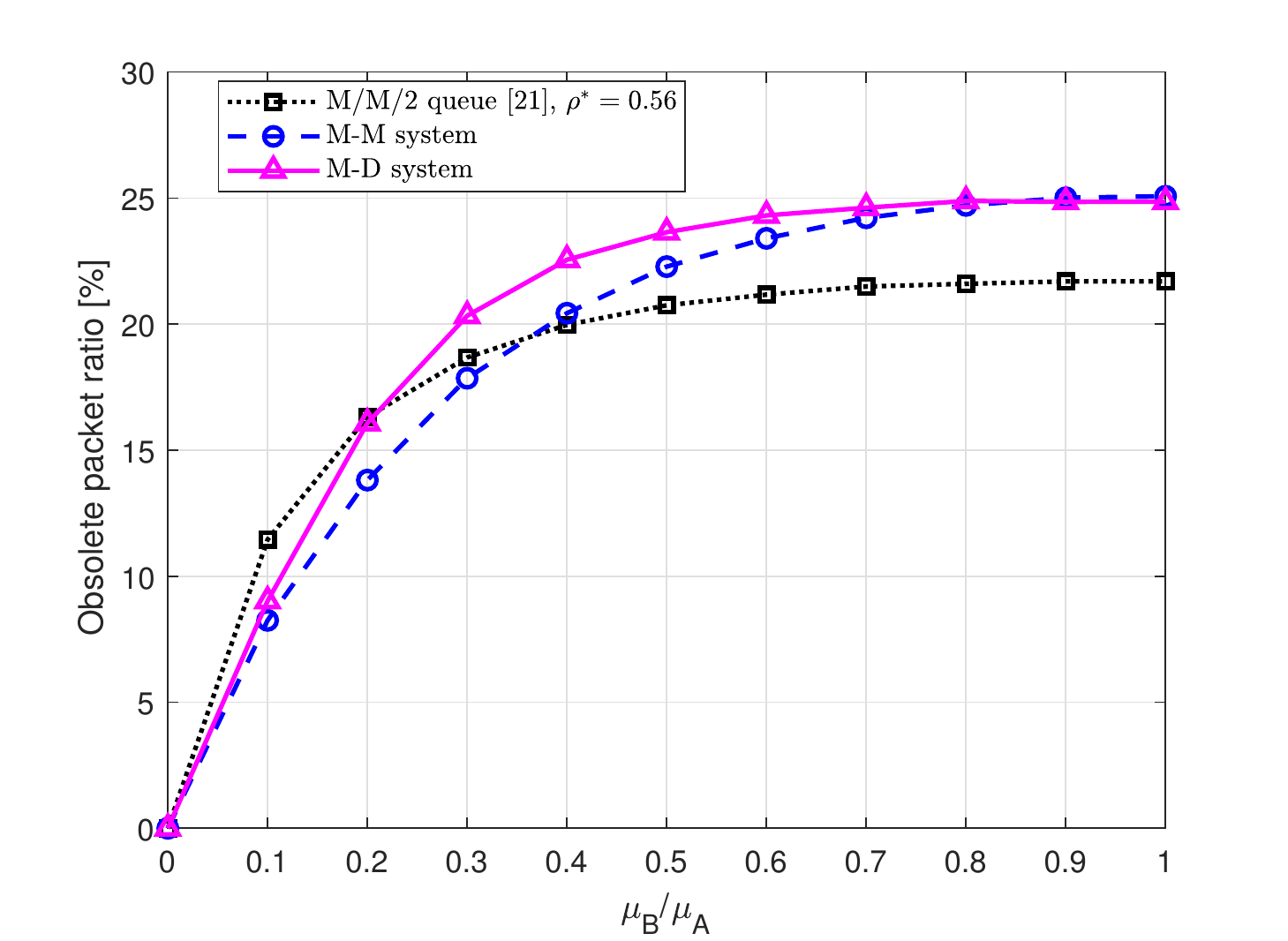}}\hspace{0.05in}
  \caption{The effective arrive rate and obsolete packet ratio of the dual-queue system as the ratio of service rate  $\mu_B/\mu_A$ increasing from $0$ to $1$. The service rate of sensor A is set as $\mu_A=1$.}
  \label{Fig_lose_arrive_rate}
  \vspace{-1.5em}
\end{figure}

Fig.~\ref{fig_effective_rate} plots the trend of the effective arrival rate of the  M-M system, M-D system, and M/M/2 queue as the ratio of service rate $\mu_{B}/\mu_{A}$ increases from 0 to 1, where the service rate of sensor A is fixed as $\mu_{A}=1$. As we can see that the effective arrival rate of the above three systems increases with an increase in the service rate ratio $\mu_{B}/\mu_{A}$. The M/M/2 queue has the lowest effective arrival rate and the worst AoI performance among the three systems. It is worth noting that when $\mu_A/\mu_B=1$, the effective arrival rates of both the M-M system and M-D system are 1.5, i.e., the $75\%$ of the total service rate $\mu=\mu_A+\mu_B=2$.
Fig.~\ref{fig_expired_rate}  depicts the obsolete packet  ratio of the  M-M system, M-D system, and M/M/2 queue as a function of  the ratio of service rate $\mu_{B}/\mu_{A}$.
From this figure, we can see that as the ratio of service rate $\mu_A/\mu_B$ increases, the system can transmit more updates, resulting  in a decrease in AoI. However, the percentage of obsolete packets in the total transmitted packets is also increasing, i.e., the resource waste rate of the considered system is increasing.
Note that when $\mu_A/\mu_B=1$, the obsolete packet  ratios of the  M-M system and M-D system both approach $25\%$, which agrees with the phenomenon observed in Fig.~\ref{fig_effective_rate}.

\begin{figure}[t]
  \centering
\subfigure[Average PAoI]{\label{fig_pree_PAoI}\includegraphics[scale=0.54]{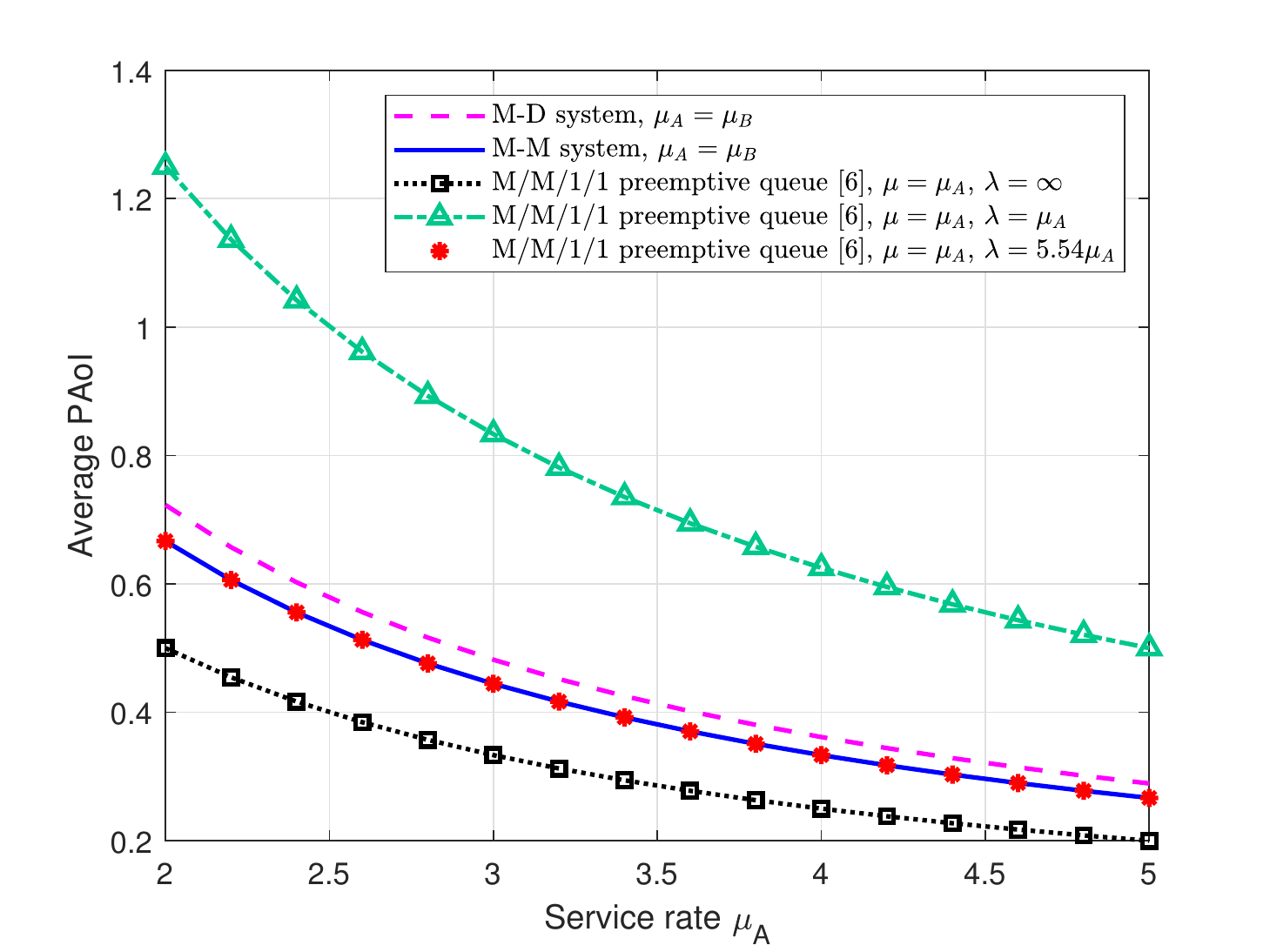}}
\subfigure[Average AoI]{\label{fig_pree_AoI}\includegraphics[scale=0.54]{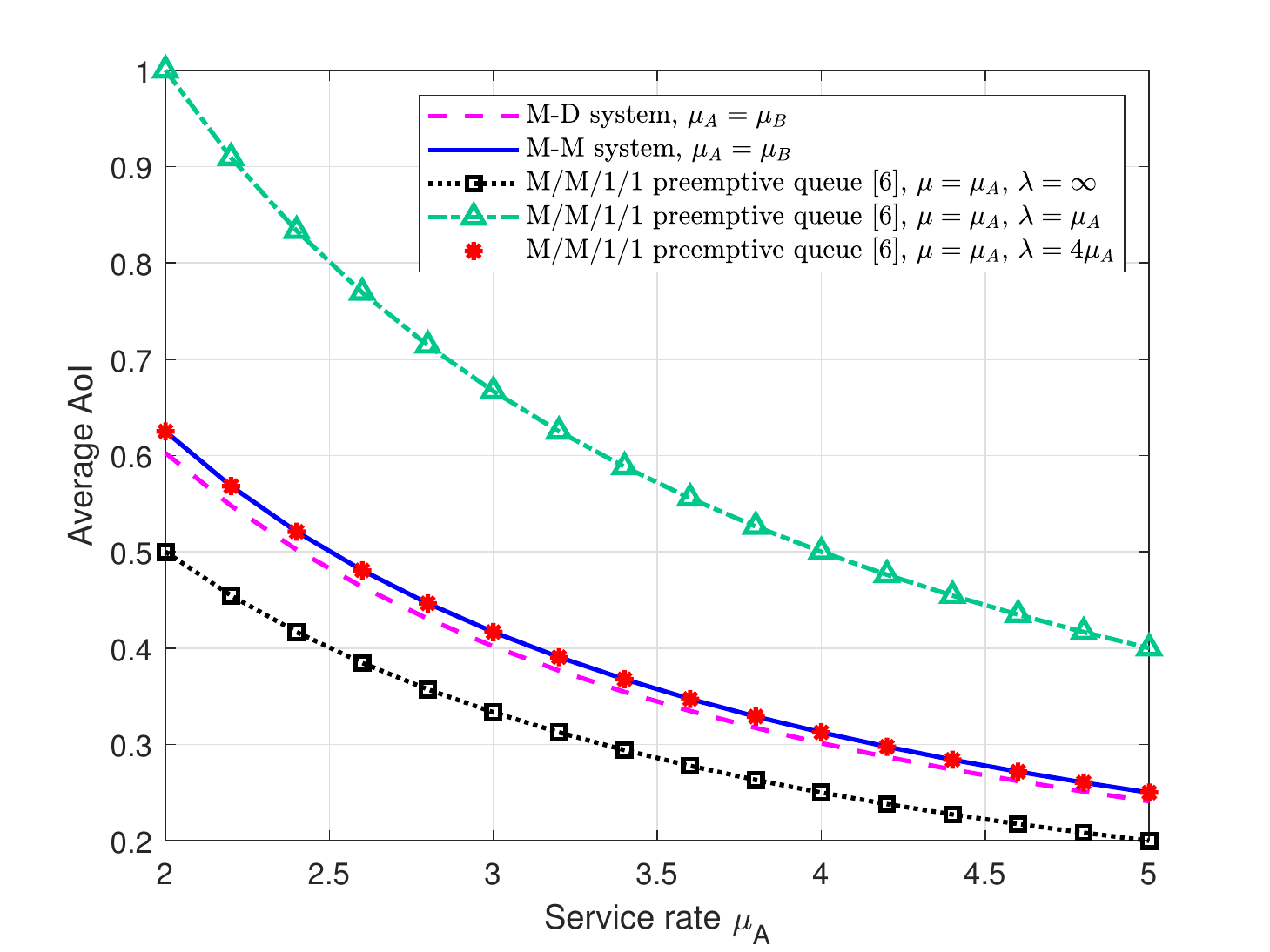}}\hspace{0.05in}
  \caption{Comparison of the AoI performance between dual-queue systems and the M/M/1/1 preemptive queue.}
  \label{Fig_preemptive_compare}
  \vspace{-1.5em}
\end{figure}

In Fig.~\ref{Fig_preemptive_compare}, we compare the AoI performance of dual-queue systems with the single-stream M/M/1/1 preemptive system (see \cite{ut_1}) under the condition that each server has the same service rate.
As we can see from Fig.~\ref{fig_pree_PAoI}, in  terms of average PAoI performance, the M-M system is equivalent to an M/M/1/1 preemptive queue with generation rate $\lambda=5.54\mu$. Fig.~\ref{fig_pree_AoI} shows that the average AoI performance of the M-M system is the same as the M/M/1/1 preemptive queue with generation rate $\lambda=4\mu$.

\section{Conclusion}\label{Conclusion}
This paper conducted an analytical study to the AoI performance of a dual-queue real-time monitoring system, in which two sensors observe the same physical process.
We considered two different multi-queue systems, namely the M-M system where the service time of the two servers is exponentially distributed and the M-D system  in which the service time of one server is exponentially distributed, and that of the other is deterministic.
We derived the closed-form expressions of the average PAoI and average AoI for both systems.
Our analysis reveals that for a single-queue update system with an exponentially distributed service time, adding a dual sensor can reduce the average PAoI by $33.3\%$ and the average AoI by $37.5\%$. Adding a sensor with the same service rate and fixed service time can reduce the average PAoI by $27.7\%$ and the average AoI by $39.7\%$.
% Compared with the single-source just-in-time update system with exponential service time distribution, the average PAoI of the  M-M system is reduced by $33.3\%$, the average AoI is reduced by $37.5\%$; The average PAoI of the  M-D system is reduced by $27.7\%$, and the average AoI is reduced by $39.7\%$.
In addition,  when the service rates of the two sensors are equal, the M-M system behaves better than the M-D system in terms of the average PAoI.  In contrast,  the M-D system outperforms the M-M system in terms of the average AoI. The numerical results showed that the multi-queue parallel transmission method can effectively improve the freshness of received information on the monitor.

\appendix
\allowdisplaybreaks
\subsection{Proof of Lemma \ref{lem_X}}\label{prf:lem_x}
First, let
\begin{align}
  G=\int_{0}^{T-x_{2}-x{1}}\cdots\int_{0}^{T-\sum_{i=1}^{n-1}x_{i}}1{\rm d}x_{n}\cdots {\rm d}x_{3}.
\end{align}
For clarity, we define $C_{r}=T-\sum_{i=1}^{r}x_{i}$ for $r=2,\cdots,n-1$,  then $G$ is written as
\begin{align}
  G&=\int_{0}^{C_{2}}\cdots\int_{0}^{C_{n-1}}1{\rm d}x_{n}\cdots {\rm d}x_{3}
   =\int_{0}^{C_{2}}\cdots\int_{0}^{C_{n-2}}C_{n-1}{\rm d}x_{n-1}\cdots {\rm d}x_{3}\nonumber\\
  &=\int_{0}^{C_{2}}\cdots\int_{0}^{C_{n-3}}\frac{C_{n-2}^{2}}{2}{\rm d}x_{n-2}\cdots {\rm d}x_{3}
  =\int_{0}^{C_{2}}\!\cdots\!\int_{0}^{C_{n-3}}\frac{(C_{n-3}-x_{n-2})^2}{2}{\rm d}x_{n-2}\cdots {\rm d}x_{3}.
\end{align}
By further denoting $h_{n-3}=C_{n-3}-x_{n-2}$, it  yields that
\begin{align}
G&=\int_{0}^{C_{2}}\!\cdots\!\int_{0}^{C_{n-4}}\int_{0}^{C_{n-3}}\frac{(h_{n-3})^2}{2}{\rm d}h_{n-3}{\rm d}x_{n-3}\cdots {\rm d}x_{3}
=\int_{0}^{C_{2}}\!\cdots\!\int_{0}^{C_{n-4}}\frac{C_{n-3}^{3}}{2 \times 3}{\rm d}x_{n-3}\cdots {\rm d}x_{3}\nonumber\\
&=\int_{0}^{C_{2}}\!\cdots\!\int_{0}^{C_{n-4}}\frac{(C_{n-4}-x_{n-3})^{3}}{2 \times 3}{\rm d}x_{n-3}\cdots {\rm d}x_{3}.
\end{align}
Similarly, let $h_{n-4}=C_{n-4}-x_{n-3}$, it results in
\begin{align}
G\!=\!\int_{0}^{C_{2}}\!\cdots\!\int_{0}^{C_{n-4}}\frac{(h_{n-4})^{3}}{2 \!\times\! 3}{\rm d}h_{n-4}\cdots {\rm d}x_{3}
\!=\!\int_{0}^{C_{2}}\!\cdots\!\int_{0}^{C_{n-5}}\!\frac{(C_{n-4})^{4}}{2\! \times\! 3\!\times\! 4}{\rm d}x_{n-4}\cdots {\rm d}x_{3}.
\end{align}
By conducting the same method, we can get that
\begin{equation}
  G=\frac{C_{2}^{n-2}}{(n-2)!}=\frac{(T-x_{2}-x_{1})^{n-2}}{(n-2)!}.
\end{equation}
This completes the proof of Lemma \ref{lem_X}.

\ifCLASSOPTIONcaptionsoff
  \newpage
\fi

\bibliography{IEEEabrv,mybib}

\begin{thebibliography}{10}
\providecommand{\url}[1]{#1}
\csname url@samestyle\endcsname
\providecommand{\newblock}{\relax}
\providecommand{\bibinfo}[2]{#2}
\providecommand{\BIBentrySTDinterwordspacing}{\spaceskip=0pt\relax}
\providecommand{\BIBentryALTinterwordstretchfactor}{4}
\providecommand{\BIBentryALTinterwordspacing}{\spaceskip=\fontdimen2\font plus
\BIBentryALTinterwordstretchfactor\fontdimen3\font minus
  \fontdimen4\font\relax}
\providecommand{\BIBforeignlanguage}[2]{{%
\expandafter\ifx\csname l@#1\endcsname\relax
\typeout{** WARNING: IEEEtran.bst: No hyphenation pattern has been}%
\typeout{** loaded for the language `#1'. Using the pattern for}%
\typeout{** the default language instead.}%
\else
\language=\csname l@#1\endcsname
\fi
#2}}
\providecommand{\BIBdecl}{\relax}
\BIBdecl

\bibitem{AuD}
Y.~Dong, Z.~Chen, S.~Liu, P.~Fan, and K.~B. Letaief, ``Age-upon-decisions
  minimizing scheduling in {Internet of Things}: To be random or to be
  deterministic?'' \emph{{IEEE} Internet Things J.}, vol.~7, no.~2, pp.
  1081--1097, Feb. 2020.

\bibitem{pappas_mag}
M.~A. Abd-Elmagid, N.~Pappas, and H.~S. Dhillon, ``On the role of age of
  information in the internet of things,'' \emph{IEEE Commun. Mag.}, vol.~57,
  no.~12, pp. 72--77, Dec. 2019.

\bibitem{wbsn}
S.-L. Chen, H.-Y. Lee, C.-A. Chen, H.-Y. Huang, and C.-H. Luo, ``Wireless body
  sensor network with adaptive low-power design for biometrics and healthcare
  applications,'' \emph{{IEEE} Syst. J.}, vol.~3, no.~4, pp. 398--409, Dec.
  2009.

\bibitem{vehicle_2}
Z.~Jiang, S.~Fu, S.~Zhou, Z.~Niu, S.~Zhang, and S.~Xu, ``{AI}-assisted low
  information latency wireless networking,'' \emph{{IEEE} Wireless Commun.},
  vol.~27, no.~1, pp. 108--115, Feb. 2020.

\bibitem{01}
S.~{Kaul}, M.~{Gruteser}, V.~{Rai}, and J.~{Kenney}, ``Minimizing age of
  information in vehicular networks,'' in \emph{Proc. 8th Annu. IEEE
  Commun.Soc. Conf. Sensor, Mesh Ad-Hoc Commun. Netw. (SECON)}, June 2011, pp.
  350--358.

\bibitem{ut_1}
S.~K. Kaul, R.~D. Yates, and M.~Gruteser, ``Status updates through queues,'' in
  \emph{Proc. 46th Annu. Conf. Inf. Sci. Syst.}, Mar. 2012, pp. 1--6.

\bibitem{ut_2}
S.~Kaul, R.~Yates, and M.~Gruteser, ``Real-time status: How often should one
  update?'' in \emph{Proc. IEEE INFOCOM}, Mar. 2012, pp. 2731--2735.

\bibitem{ut_4}
R.~D. Yates, Y.~Sun, D.~R. Brown, S.~K. Kaul, E.~Modiano, and S.~Ulukus, ``Age
  of information: An introduction and survey,'' \emph{{IEEE} J. Sel. Areas
  Commun.}, vol.~39, no.~5, pp. 1183--1210, May 2021.

\bibitem{book}
A.~Kosta, N.~Pappas, and V.~Angelakis, ``Age of information: A new concept,
  metric, and tool,'' \emph{Foundations and Trends in Networking}, vol.~12,
  no.~3, pp. 162--259, 2017.

\bibitem{PAoI_R2}
L.~Huang and E.~Modiano, ``Optimizing age-of-information in a multi-class
  queueing system,'' in \emph{Proc. IEEE Int. Symp. Inf. Theory (ISIT)}, Jun.
  2015, pp. 1681--1685.

\bibitem{ref_jit}
M.~Costa, M.~Codreanu, and A.~Ephremides, ``On the age of information in status
  update systems with packet management,'' \emph{IEEE Trans. Inf. Theory},
  vol.~62, no.~4, pp. 1897--1910, Apr. 2016.

\bibitem{A_general}
Y.~Inoue, H.~Masuyama, T.~Takine, and T.~Tanaka, ``A general formula for the
  stationary distribution of the age of information and its application to
  single-server queues,'' \emph{IEEE Trans. Inf. Theory}, vol.~65, no.~12, pp.
  8305--8324, Aug. 2019.

\bibitem{pappas_icc}
N.~Pappas, J.~Gunnarsson, L.~Kratz, M.~Kountouris, and V.~Angelakis, ``Age of
  information of multiple sources with queue management,'' in \emph{Proc. IEEE
  Int. Conf. Commun. (ICC)}, Jun. 2015, pp. 5935--5940.

\bibitem{policy_1}
E.~Najm and E.~Telatar, ``Status updates in a multi-stream {M/G/1/1} preemptive
  queue,'' in \emph{Proc. IEEE Conf. Comput. Commun. (INFOCOM) Workshops}, Apr.
  2018, pp. 124--129.

\bibitem{my_tcom}
Z.~Chen, D.~Deng, C.~She, Y.~Jia, L.~Liang, S.~Fang, M.~Wang, and Y.~Li, ``Age
  of information: The multi-stream {M/G/1/1} non-preemptive system,''
  \emph{{IEEE} Trans. Commun.}, pp. 1--1, 2022.

\bibitem{pappas_iotj}
G.~Stamatakis, N.~Pappas, and A.~Traganitis, ``Optimal policies for status
  update generation in an {IoT} device with heterogeneous traffic,''
  \emph{{IEEE} Internet Things J.}, vol.~7, no.~6, pp. 5315--5328, Jun. 2020.

\bibitem{multi_path}
J.~Doncel and M.~Assaad, ``Age of information of jackson networks with finite
  buffer size,'' \emph{IEEE Wireless Commun. Lett.}, vol.~10, no.~4, pp.
  902--906, Apr. 2021.

\bibitem{multi_server}
A.~M. Bedewy, Y.~Sun, and N.~B. Shroff, ``Optimizing data freshness,
  throughput, and delay in multi-server information-update systems,'' in
  \emph{Proc. IEEE Int. Symp. Inf. Theory (ISIT)}, Jul. 2016, pp. 2569--2573.

\bibitem{parallel_MM11}
J.~Doncel and M.~Assaad, ``Age of information in a decentralized network of
  parallel queues with routing and packets losses,'' \emph{Journal of
  Communications and Networks}, vol.~24, no.~1, pp. 17--20, Feb. 2022.

\bibitem{GG}
Y.~Inoue, ``The probability distribution of the {AoI} in queues with infinitely
  many servers,'' in \emph{Proc. IEEE INFOCOM Conf. Comput. Commun. Workshops
  (INFOCOM WKSHPS}, Jul. 2020, pp. 297--302.

\bibitem{MM2_conference}
C.~Kam, S.~Kompella, and A.~Ephremides, ``Effect of message transmission
  diversity on status age,'' in \emph{Proc. IEEE Int. Symp. Inf. Theory
  (ISIT)}, Jun. 2014, pp. 2411--2415.

\bibitem{MM2}
C.~Kam, S.~Kompella, G.~D. Nguyen, and A.~Ephremides, ``Effect of message
  transmission path diversity on status age,'' \emph{IEEE Trans. Inf. Theory},
  vol.~62, no.~3, pp. 1360--1374, Mar. 2016.

\bibitem{MPTCP}
K.~Gao, C.~Xu, X.~Ji, J.~Qin, S.~Yang, L.~Zhong, and D.~Wu, ``Freshness-aware
  age optimization for multipath {TCP} over software defined networks,''
  \emph{IEEE Trans. Netw. Sci. Eng.}, pp. 1--12, Apr. 2021.

\bibitem{parallel_c}
R.~D. Yates, ``Status updates through networks of parallel servers,'' in
  \emph{Proc. IEEE Int. Symp. Inf. Theory (ISIT)}, Jun. 2018, pp. 2281--2285.

\bibitem{closed_form}
Z.~Zhang, X.~Zhu, Y.~Jiang, J.~Cao, and Y.~Liu, ``Closed-form {AoI} analysis
  for dual-queue short-block transmission with block error,'' in \emph{Proc.
  IEEE Wireless Commun. Netw. Conf. (WCNC)}, Mar. 2021, pp. 1--6.

\bibitem{dual}
X.~Jia, S.~Cao, and M.~Xie, ``Age of information of dual-sensor information
  update system with {HARQ} chase combining and energy harvesting diversity,''
  \emph{IEEE Wireless Commun. Lett.}, vol.~10, no.~9, pp. 2027--2031, Sep.
  2021.

\bibitem{Related_02}
J.~Hribar, M.~Costa, N.~Kaminski, and L.~A. DaSilva, ``Using correlated
  information to extend device lifetime,'' \emph{{IEEE} Internet Things J.},
  vol.~6, no.~2, pp. 2439--2448, Apr. 2019.

\bibitem{camera}
Q.~He, G.~D\'{a}n, and V.~Fodor, ``Joint assignment and scheduling for
  minimizing age of correlated information,'' \emph{IIEEE/ACM Trans. Netw.},
  vol.~27, no.~5, pp. 1887--1900, Oct. 2019.

\bibitem{NB_IOT}
Y.-P.~E. Wang, X.~Lin, A.~Adhikary, A.~Grovlen, Y.~Sui, Y.~Blankenship,
  J.~Bergman, and H.~S. Razaghi, ``A primer on {3GPP} narrowband {Internet of
  Things},'' \emph{IEEE Commun. Mag.}, vol.~55, no.~3, pp. 117--123, Mar. 2017.

\end{thebibliography}

\end{document}